\documentclass[twocolumn]{jpsj2}

\def\runauthor{}

\title{%
Electron Correlation and Pairing States in Superconductors 
without Inversion Symmetry
}

\author{%
Satoshi Fujimoto\thanks{E-mail address: fuji@scphys.kyoto-u.ac.jp}
}

\inst{%
Department of Physics, Kyoto University, Kyoto 606-8502, Japan
}

\recdate{\today}

\abst{%
This article is a pedagogical review of 
theoretical studies of noncentrosymmetric superconductors 
with particular emphasis on the role played by electron correlation, which
is important for heavy fermion systems. 
We survey unique properties of parity-violated superconductivity
such as the admixture of spin singlet and triplet states,
unusual paramagnetism, large Pauli limiting fields, magnetoelectric effects,
the helical vortex phase, and the anomalous Hall effect.
It is pointed out that these remarkable features 
are strengthened by a
strong electron correlation effect, and thus are prominent
in heavy fermion superconductors without inversion symmetry.
We also discuss possible pairing states realized in the heavy fermion 
system CePt$_3$Si. 
}

\kword{ superconductivity, heavy fermion, inversion symmetry, 
electron correlation
}

\begin{document}
\maketitle

\section{Introduction}

In all superconductors, the symmetries of systems such 
as time-reversal symmetry and inversion symmetry 
impose important constraint on pairing states.
For instance, when a system has an inversion center, Cooper pairs are classified
into a spin-singlet state or a spin-triplet state according to parity.
Conversely, in the case without inversion symmetry, in which 
an asymmetric potential gradient yields 
a spin-orbit (SO) interaction that breaks parity, 
the admixture of a spin singlet state and a spin triplet state 
is possible, and also, various exotic electromagnetic properties 
may result.
Such parity-violated superconductivity is the main topic of this article.
Pioneering theoretical studies on this subject
were initiated by Bulaevskii et al. decades ago, and developed by
Edelstein in 1989.\cite{ede,ede2,ede3,old}
Edelstein pointed out
that, in addition to the singlet-triplet mixing of pairing states, 
the parity-breaking SO interaction also gives rise to
unusual electromagnetism such as
van-Vleck-like spin susceptibility analogous to
van-Vleck orbital susceptibility, and
magnetoelectric effects, i.e., supercurrent induced by the Zeeman field,
and conversely, magnetization induced by supercurrent flow.
The latter effects are due to the nontrivial coupling between
the spin and charge degrees of freedom raised by a parity-breaking
SO interaction.
These intriguing phenomena were also investigated in more detail by
Gor'kov and Rashba, and Yip.\cite{gr,yip}
However, experimental tests on these interesting theoretical predictions
in real materials have not been performed for decades since 
the first prediction made by Edelstein.\cite{ede}

In 2004, Bauer et al. discovered
a bulk superconductor without inversion symmetry,
CePt$_3$Si,\cite{bau} and stimulated 
more extensive experimental 
and theoretical studies on this 
subject.\cite{take,neu,yasu,yogi,yogi2,iza,pene,fri,sam2,sam,ser,
kau,yip2,sam1,fuji2,haya,sem,tana,ichi,fuji3}
In particular, the ETH group and Agterberg 
revealed some novel aspects of
electromagnetic properties such as large Pauli limiting fields, and 
a possible realization of the helical
vortex state in magnetic fields.\cite{fri,kau} 
The helical vortex state is a kind of the Fulde-Ferrel state in which
Cooper pairs have a  center of mass momentum, and
the phase of the order parameter is spatially modulated.
This Fulde-Ferrel-like state was also independently predicted by
Samokhin on the basis of the general GL theory with strong SO
coupling.\cite{sam2}

After this breakthrough, other novel heavy fermion
superconductors without inversion symmetry such as 
UIr, CeRhSi$_3$, and CeIrSi$_3$
have been discovered.\cite{uir,kim,onuki}
Also, in non-$f$-electron systems,  
new noncentrosymmetric superconductors 
such as Cd$_2$Re$_2$O$_7$, Li$_2$Pd$_3$B, and 
Li$_2$Pt$_3$B have been discovered.\cite{cd,tog1,tog2,yu}
An intriguing feature of these systems is that according to
LDA calculations and de Haas-van Alphen experiments,
the SO splitting of the Fermi surfaces is much larger
than the superconducting gap.\cite{sam,dh,kim2,pic}
For instance, in the case of CePt$_3$Si, 
the magnitude of the SO splitting is nearly equal to 
10 percent of the Fermi energy.\cite{sam,dh}
This large SO splitting indicates the important role of parity violation
in these superconductors.

In this article, we survey theoretical studies of noncentrosymmetric
superconductors with particular emphasis on the role played by electron
correlation effects and the application to heavy fermion systems.
In heavy fermion systems where $f$-electrons are itinerant,
there exists a strong spherically-symmetric SO interaction
in addition to the parity-breaking SO interaction.
Thus, we need to clarify the relation between these different types
of SO interaction.
The parity-breaking SO interaction due to an asymmetric potential gradient
$\nabla V$ is generally expressed as
$(\mbox{\boldmath $k$}\times\nabla V)\cdot\mbox{\boldmath $\sigma$}$,
where $\mbox{\boldmath $k$}$ is the momentum, and $\mbox{\boldmath $\sigma$}$
is the Pauli matrix. This SO interaction is expressed in terms of
electron spins.
In the case of CePt$_3$Si, because of the strong symmetric SO
interaction, the ground state of the $f$-electron level
is the $\Gamma_7$ Kramers doublet, the basis of which
is,\cite{neu}
\begin{eqnarray}
|a\rangle &=&\sqrt{5/6}|5/2\rangle 
-\sqrt{1/6}|-3/2\rangle, \nonumber \\
|\bar{a}\rangle &=&\sqrt{5/6}|-5/2\rangle 
-\sqrt{1/6}|3/2\rangle. 
\end{eqnarray} 
Transforming the basis of spin to the basis $|a\rangle$, $|\bar{a}\rangle$,
we find that 
the parity-breaking SO interaction 
is rewritten as $(5/12)(\mbox{\boldmath $k$}\times\nabla V)
\cdot\mbox{\boldmath $\sigma$}$, where the Pauli matrix 
$\mbox{\boldmath $\sigma$}$ is defined in the pseudospin space spanned by 
$|a\rangle$, $|\bar{a}\rangle$.
That is, 
the matrix structure is not affected, and only the constant prefactor
is changed.
Thus, even in the case of  heavy fermion systems,
as long as the Bloch states of $f$-electrons are labeled 
by pseudospins that constitute the Kramers doublet,
we can use the same expression of the parity-breaking SO interaction
as that for electron spins.
Throughout this paper, when we consider heavy fermion systems,
we use the terms ``spin singlet state'' 
and ``spin triplet state''
to indicate the pseudospin singlet state and the pseudospin
triplet state, respectively.
In the case that upper levels of $f$-electrons are not negligible,
we need a more elaborate treatment of the parity-breaking SO interaction.

The organization of this paper is as follows:
In \S 2, we present the microscopic description
of the superconducting state without inversion
symmetry which takes electron correlation effects
into account.
This section is a bit technical.
Readers who are not interested in the technical details 
may proceed to the next sections.
In \S 3, we argue the pairing state 
on the basis of the BCS weak-coupling
approximation, focusing on the admixture of spin singlet and triplet states.
In \S 4, we study unique electromagnetic properties of
noncentrosymmetric superconductors.
In \S 5, we provide a discussion on possible pairing states
realized in CePt$_3$Si.
In \S 6, we give concluding remarks.

\section{Fermi Liquid Theory for Superconductors 
without Inversion Symmetry}

In this section, we present the basic formulation
for the BCS theory of noncentrosymmetric superconductors,
which includes electron correlation effects as the Fermi 
liquid corrections.\cite{fuji2,fuji3}
We are mainly 
concerned with electromagnetic properties.
As is shown below, 
unique features of magnetism in inversion-symmetry-broken systems appear
as a result of paramagnetic effects rather than diamagnetic effects.
Thus we take into account the Zeeman coupling with an
external magnetic field $\mbox{\boldmath $H$}=(H_x,H_y,H_z)$,
neglecting the orbital diamagnetic effect for a while.
Thus, our analysis is based on the following simple Hamiltonian, 
\begin{eqnarray}
\mathcal{H}&=&\mathcal{H}_0+\mathcal{H}_{\rm SO}
+\mathcal{H}_{\rm Zeeman}, 
\label{ham} \\
\mathcal{H}_0&=&\sum_{k,\sigma} \varepsilon_k c^{\dagger}_{k}c_{k}
+U\sum_i n_{\uparrow i}n_{\downarrow i}, \label{ham0} \\
\mathcal{H}_{\rm SO}&=&\alpha\sum_{k}
c^{\dagger}_{k}\mbox{\boldmath $\mathcal{L}$}_0(k)
\cdot\mbox{\boldmath $\sigma$}c_{k},
\label{so} \\
\mathcal{H}_{\rm Zeeman}&=&\sum_kc^{\dagger}_k
\mu_{\rm B}\mbox{\boldmath $\sigma$}\cdot\mbox{\boldmath $H$}
c_k. \label{zee}
\end{eqnarray}
where $c^{\dagger}_k=(c^{\dagger}_{\uparrow k},
c^{\dagger}_{\downarrow k})$ 
is the two-component
spinor field for an electron with the spin $\uparrow$, $\downarrow$,
and the momentum $k$.
$n_{\sigma i}=c^{\dagger}_{\sigma i}c_{\sigma i}$ is 
the number density operator at the site $i$. 
The components of $\mbox{\boldmath $\sigma$}=(\sigma_x,\sigma_y,\sigma_z)$ 
are the Pauli matrix.
The vector $\mbox{\boldmath $\mathcal{L}$}_0(k)
=({\cal L}_{0x},{\cal L}_{0y},{\cal L}_{0z})$
in the spin-orbit interaction term $\mathcal{H}_{\rm SO}$
obeys $\mbox{\boldmath $\mathcal{L}$}_0(-k)
=-\mbox{\boldmath $\mathcal{L}$}_0(k)$, breaking inversion symmetry.
The explicit form of $\mbox{\boldmath $\mathcal{L}$}_0(k)$ is determined 
by details of the crystal structure.
In the case of the tetragonal structure with an asymmetric potential
gradient along the $z$-axis, $\mbox{\boldmath $\mathcal{L}$}_0(k)=(k_y,-k_x,0)$
for a small $k$, which is the Rashba interaction.\cite{ras}
In the case of cubic structures with the point group symmetry $T$, such as
the zinc-blende, 
$\mbox{\boldmath $\mathcal{L}$}_{0}(k)=(k_x(k_y^2-k_z^2),
k_y(k_z^2-k_x^2),k_z(k_x^2-k_y^2))$, which is related to the Dresselhaus
interaction.\cite{dres,dyp,sil}
For simplicity, we assume that the $g$-factor is equal to 2, and spin $s=1/2$.
As a matter of fact, for heavy fermion systems
the $g$-factor may generally be different from 2 because of
the strong symmetric SO interaction.

Electron correlation effects, which are important for heavy fermion systems,
are incorporated by on-site Coulomb repulsion $U$.
We also assume that, in addition to the on-site repulsion,  
there is an effective pairing interaction with an angular momentum 
$\ell\geq 1$,
which may stem from the on-site Coulomb interaction in (\ref{ham}) 
or from any other factors
not included in the Hamiltonian (\ref{ham}).

To calculate physical quantities microscopically,
we utilize the single-particle Green function, the inverse 
of which is, in the conventional Nambu representation,
defined as
\begin{eqnarray}
\hat{\mathcal{G}}^{-1}(p)=
\left(
\begin{array}{cc}
   i\varepsilon_n-\hat{H}(p) & -\hat{\Delta}(p)   \\
   -\hat{\Delta}^{\dagger}(p) & i\varepsilon_n+\hat{H}^{t}(-p)
\end{array}
\right), \label{ginsc}
\end{eqnarray}
where $p=(i\varepsilon_n,\mbox{\boldmath $k$})$ and
\begin{eqnarray}
\hat{H}(p)=\varepsilon_k-\mu+
\alpha\mbox{\boldmath $\mathcal{L}$}_0(k)\cdot\mbox{\boldmath $\sigma$}
+\hat{\Sigma}(p)+\mu_{\rm B}\mbox{\boldmath $\sigma$}\cdot
\mbox{\boldmath $H$}, \label{h1}
\end{eqnarray}
with $\mu$ being the chemical potential. 
The self-energy matrix $\hat{\Sigma}$ consists of 
both diagonal and off-diagonal components, and expressed as
\begin{eqnarray}
\hat{\Sigma}&=&
\left(
\begin{array}{cc}
 \Sigma_{\uparrow\uparrow}(p)   & \Sigma_{\uparrow\downarrow}(p)  \\
 \Sigma_{\downarrow\uparrow}(p) & \Sigma_{\downarrow\downarrow}(p)
\end{array}
\right)   \nonumber \\
&=&\Sigma_0+\mbox{\boldmath $\Sigma$}
\cdot\mbox{\boldmath $\sigma$}, \label{self}
\end{eqnarray}
where $\mbox{\boldmath $\Sigma$}=(\Sigma_x,\Sigma_y,\Sigma_z)$ 
with $\Sigma_0=(\Sigma_{\uparrow\uparrow}+\Sigma_{\downarrow\downarrow})/2$,
$\Sigma_x=(\Sigma_{\downarrow\uparrow}+\Sigma_{\uparrow\downarrow})/2$, 
$\Sigma_y=(\Sigma_{\downarrow\uparrow}-\Sigma_{\uparrow\downarrow})/2i$, and 
$\Sigma_z=(\Sigma_{\uparrow\uparrow}-\Sigma_{\downarrow\downarrow})/2$.

The SO interaction $\alpha \mbox{\boldmath $\mathcal{L}$}_0(k)
\cdot\mbox{\boldmath $\sigma$}$ splits the Fermi surface into
two parts as shown in Fig. \ref{sos}.
The SO splitted two bands are derived by diagonalizing 
$i\varepsilon_n-\hat{H}(p)$ and $i\varepsilon_n+\hat{H}^t(-p)$ in 
$\hat{\mathcal{G}}^{-1}(p)$.
The diagonalization is a bit tricky because
$\mbox{\boldmath $\Sigma$}(p)$ is generally non-Hermitian.
In the absence of the magnetic field,
we can verify that $\Sigma_x(p)$ and $\Sigma_y(p)$ are 
real quantities, and thus $\mbox{\boldmath $\Sigma$}(p)$ is 
Hermitian.\cite{fuji3}
Also, as long as the linear response against the magnetic field is concerned,
the imaginary part of 
$\Sigma_x(p)$ and $\Sigma_y(p)$
are negligible compared to the real parts even for 
$\mbox{\boldmath $H$}\neq 0$.
Then, the diagonalization 
is achieved by using the unitary transformation 
$\hat{\mathcal{A}}(p)\hat{\mathcal{G}}^{-1}(p)\hat{\mathcal{A}}^{\dagger}(p)$
with
\begin{eqnarray}
\hat{\mathcal{A}}(p)=
\left(
\begin{array}{cc}
\hat{U}(p) & 0 \\
 0 & \hat{U}^{t\dagger}(-p)
\end{array}
\right),
\end{eqnarray}
\begin{eqnarray}
\hat{U}(p)=
\left(
\begin{array}{cc}
\xi_{+}(p)  &   \xi_{-}(p)\eta_{-}(p) \\
-\xi_{-}(p)\eta_{+}(p) & \xi_{+}(p) 
\end{array}
\right), \label{unita}
\end{eqnarray}
\begin{eqnarray}
\xi_{\pm}(p)=\frac{1}{\sqrt{2}}\left[1\pm\frac{{\cal L}_z(p)}
{\Vert\mbox{\boldmath $\mathcal{L}$}(p)\Vert}
\right]^{\frac{1}{2}},
\end{eqnarray}
\begin{eqnarray}
\eta_{\pm}(p)=\frac{{\cal L}_x(p)\pm i{\cal L}_y(p)}
{\sqrt{{{\cal L}_x(p)}^2+{{\cal L}_y(p)}^2}}.
\end{eqnarray}
Here, 
\begin{eqnarray}
\mbox{\boldmath $\mathcal{L}$}(p)=({\cal L}_x(p),{\cal L}_y(p),{\cal L}_z(p))
=\mbox{\boldmath $\mathcal{L}$}_{0}(k)
-\frac{\mu_{\rm B}}{\alpha}\mbox{\boldmath $H$}+\frac{1}{\alpha}
\mbox{\boldmath $\Sigma$}(p).
\end{eqnarray}
and $\Vert\mbox{\boldmath $\mathcal{L}$}(p)\Vert=
\sqrt{\mathcal{L}_x^2+\mathcal{L}_y^2+\mathcal{L}_z^2}$.

In the absence of the magnetic field $\mbox{\boldmath $H$}=0$,
the superconducting 
gap function $\hat{\Delta}(p)$ in eq. (\ref{ginsc})
is also diagonalized by $\hat{\mathcal{A}}$ if and only if
the gap function has the form,\cite{ede,fri,fuji2}
\begin{eqnarray}
\hat{\Delta}(p)=\Delta_s(k)i\sigma_y+\Delta_t(k)\mbox{\boldmath 
$\mathcal{L}$}(p)\cdot\mbox{\boldmath $\sigma$}i\sigma_y \label{delt}
\end{eqnarray}
That is, when the $\mbox{\boldmath $d$}$-vector of the triplet gap 
satisfies $\mbox{\boldmath $d$}\propto\mbox{\boldmath 
$\mathcal{L}$}(p)$, 
there is no Cooper pair between electrons on
the different Fermi surfaces splitted by the SO interaction.
Conversely, unless $\mbox{\boldmath $d$}\propto\mbox{\boldmath 
$\mathcal{L}$}(p)$,
pairing correlation between the SO splitted two bands is induced,
which yields pair-breaking effects. (see Fig.\ref{sos}.)
According to weak-coupling calculations performed by 
Frigeri et al.\cite{fri}, the $\mbox{\boldmath $d$}$-vector
parallel to $\mbox{\boldmath $\mathcal{L}$}(p)$
gives the highest $T_c$, as long as the pairing interaction
stabilizes the gap function with the same momentum dependence 
as that of $\mbox{\boldmath $\mathcal{L}$}(p)$.
In general, a finite magnetic field gives rise to pairing correlations
between the different Fermi surfaces, and thus the gap $\hat{\Delta}(p)$
can not be diagonalized by  $\hat{\mathcal{A}}$.
However, within the linear response theory, which is adequate for
the following argument, the correlation between the different Fermi
surfaces is negligible, 
and the inverse of eq.(\ref{ginsc}) is given by,
\begin{eqnarray}
\hat{\mathcal{G}}(p)=
\left(
\begin{array}{cc}
 \hat{G}(p)  & \hat{F}(p)   \\
  \hat{F}^{\dagger}(p) &  -\hat{G}^{t}(-p) 
\end{array}
\right), \label{g1}
\end{eqnarray}
where
\begin{eqnarray}
\hat{G}(p)=\sum_{\tau=\pm 1}
\frac{1+\tau\hat{\mbox{\boldmath $\mathcal{L}$}}(p)
\cdot\mbox{\boldmath $\sigma$}}{2}G_{\tau}(p),
\end{eqnarray}
\begin{eqnarray}
\hat{F}(p)=\sum_{\tau=\pm 1}
\frac{1+\tau\hat{\mbox{\boldmath $\mathcal{L}$}}(p)\cdot
\mbox{\boldmath $\sigma$}}{2}i\sigma_yF_{\tau}(p),
\label{fff}
\end{eqnarray}
and
\begin{eqnarray}
G_{\tau}(p)=\frac{z_{k\tau}(i\varepsilon+\varepsilon_{k\tau}^{*})}
{(i\varepsilon+i\gamma_{k\tau} {\rm sgn}\varepsilon)^2-E^{2}_{k\tau}},
\label{scg}
\end{eqnarray}
\begin{eqnarray}
F_{\tau}(p)=\frac{z_{k\tau}\Delta_{k\tau}}
{(i\varepsilon+i\gamma_{k\tau} {\rm sgn}\varepsilon)^2-E^{2}_{k\tau}}.
\label{scf}
\end{eqnarray}
Here, $E_{k\tau}=\sqrt{\varepsilon_{k\tau}^{*2}+\Delta^2_{k\tau}}$,
 $\Delta_{k\tau}=z_{k\tau}\tilde{\Delta}_{k\tau}$ with
\begin{eqnarray}
\tilde{\Delta}_{k\tau}=\Delta_s(k)
+\tau|\mbox{\boldmath $\mathcal{L}$}(k)|\Delta_t(k),
\label{deldia}
\end{eqnarray}
and $\varepsilon_{k\tau}^{*}$ is the quasiparticle energy determined by
using
\begin{equation}
\varepsilon_{k\tau}^{*}+\mu-\varepsilon_{k}
-\Sigma_{0}(\varepsilon_{\tau}^{*},\mbox{\boldmath $k$})
-\tau\alpha \Vert\mbox{\boldmath $\mathcal{L}$}
(\varepsilon_{k\tau}^{*},\mbox{\boldmath $k$})
\Vert=0. \label{qene}
\end{equation}
$\gamma_{k\tau}$ is the quasi-particle damping given by
\begin{eqnarray}
\gamma_{k\tau}=z_{k\tau}({\rm Im}\Sigma_0^{R}
(\varepsilon,\mbox{\boldmath $k$})
+\alpha{\rm Re}\hat{\mbox{\boldmath $\mathcal{L}$}}^R
\cdot{\rm Im}\mbox{\boldmath $\Sigma$}^R
(\varepsilon,\mbox{\boldmath $k$})
)|_{\varepsilon\rightarrow i\varepsilon_n},
\label{damp}
\end{eqnarray}
where ${\rm Re}\hat{\mbox{\boldmath $\mathcal{L}$}}^R=
{\rm Re}\mbox{\boldmath $\mathcal{L}$}^R
(\varepsilon,\mbox{\boldmath $k$})/|{\rm Re}\mbox{\boldmath $\mathcal{L}$}^R
(\varepsilon,\mbox{\boldmath $k$})|$
and quantities with the superscript $R$ indicate 
retarded functions obtained by
the analytic continuation $i\varepsilon_n\rightarrow \varepsilon+i\delta$ 
($\delta>0$).
The mass renormalization factor $z_{k\tau}$ is defined as
\begin{eqnarray}
z_{k\tau}=\biggl[1-\frac{\partial \Sigma_0(p)}
{\partial (i\varepsilon_n)} 
-\tau \alpha \frac{\partial \Vert\mbox{\boldmath $\mathcal{L}$}(p)\Vert}
{\partial (i\varepsilon_n)}
\biggr]^{-1}\biggr|_{i\varepsilon_n\rightarrow\varepsilon^{*}_{k\tau}}.
\label{zkt}
\end{eqnarray}
In the derivation of the above equations, we assumed that 
the quasiparticle approximation is applicable.
As will be seen in the following, when the SO splitting is
much larger than the superconducting gap,
some electromagnetic properties are dominated by electrons far 
from the Fermi surfaces. In such cases,
the quasiparticle approximation is not generally applicable, and 
one needs to take into account the full-energy dependence of
the self-energy.

\section{Pairing State --- 
Admixture of Spin Singlet State and Spin Triplet State}

\begin{figure}
\begin{center}
\includegraphics[width=6cm]{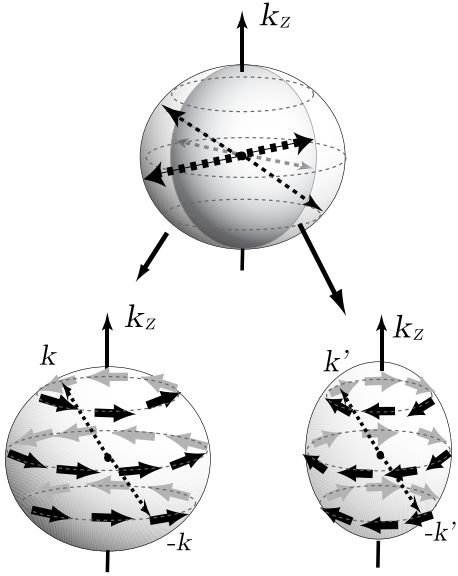}
\end{center}
\caption{
Two Fermi surfaces splitted by the Rashba SO interaction.
On each Fermi surface, spins are parallel to 
$\mbox{\boldmath $k$}\times\nabla V$, and have a definite eigen value
of $(\mbox{\boldmath $k$}\times \mbox{\boldmath $n$})\cdot
\mbox{\boldmath $\sigma$}/|\mbox{\boldmath $k$}|$; i.e. +1 or -1.
When $\mbox{\boldmath $d$}\propto \mbox{\boldmath $\mathcal{L}$}_0(k)$,
Cooper pairs are formed on each Fermi surfaces 
(black and gray dotted thin arrows).
When the $\mbox{\boldmath $d$}$-vector is not parallel to 
$\mbox{\boldmath $\mathcal{L}$}_0(k)$, the pairing correlation between 
the different Fermi surfaces exists (black dotted thick arrows), 
which gives rise to pair-breaking effects.
}
\label{sos}
\end{figure}

\begin{figure}
\begin{center}
\includegraphics[width=5cm]{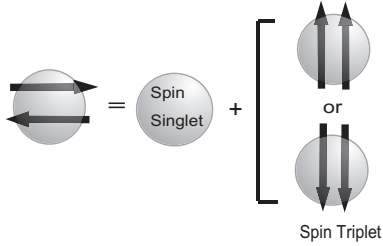}
\end{center}
\caption{Schematic picture of parity-violated Cooper pairs.
The left-hand side represents the Cooper pair 
$|\rightarrow\rangle|\leftarrow\rangle$ with the spin
quantization axis parallel to $\mbox{\boldmath $k$}\times\mbox{\boldmath $n$}$.
The right-hand side represents the sum of a spin singlet state and
a spin triplet state with $S_z=\pm 1$ for the spin quantization axis
parallel to the $z$-axis.}
\label{mix}
\end{figure}

One of the most remarkable features of noncentrosymmetric 
superconductors is 
the admixture of spin singlet and spin triplet states.\cite{ede,gr,fri}
This phenomenon is conceptually understood as follows.
When a system has a crystal structure without an inversion center,
an asymmetric potential gradient $\nabla V$ exists, causing
the spin-orbit (SO) interaction $(\mbox{\boldmath $k$}\times \nabla V)\cdot
\mbox{\boldmath $\sigma$}$, which breaks parity 
as well as spin inversion symmetry.
The asymmetric SO interaction splits the Fermi surfaces into two parts.
We show a simple example in Fig.\ref{sos} in the case of
$\nabla V \parallel \mbox{\boldmath $n$}\equiv (001)$, 
which corresponds to the Rashba-type SO interaction.
In Fig. \ref{sos}, the spin quantization axis is parallel to
$\mbox{\boldmath $k$}_F\times \nabla V$.
Then, on one of the SO splitted Fermi surfaces, a Cooper pair consists of
an electron with the momentum $k$ and spin $\rightarrow$, and an electron
with the momentum $-k$ and spin $\leftarrow$; 
i.e. $|k\rightarrow\rangle|-k\leftarrow\rangle$.
Since the counterpart of this state, 
$|\leftarrow\rangle |\rightarrow\rangle$,
is formed on the other Fermi surface, the superposition of these states
is impossible.
The pairing state $|\rightarrow\rangle|\leftarrow\rangle$
is an admixture of a spin singlet state and a spin triplet state with
an in-plane spin projection equal to zero, i.e.,
$|\rightarrow\rangle|\leftarrow\rangle =
[(|\rightarrow\rangle|\leftarrow\rangle - 
|\leftarrow\rangle|\rightarrow\rangle)
+(|\rightarrow\rangle|\leftarrow\rangle + 
|\leftarrow\rangle|\rightarrow\rangle)]/2$.
This means that for the spin quantization axis parallel to the $z$-axis,
the spin triplet component has the nonzero spin projection, $S_z=\pm 1$.
(see Fig.\ref{mix}.)
More precisely, the spin part of
the triplet pairs is determined by the parity-breaking SO interaction,
ensuring that the pairing correlation
between different Fermi surfaces, which yields
pair breaking effects, is suppressed, as pointed out in the previous section.

To obtain further insight into the pairing state, we switch to a microscopic
argument based on the BCS theory.
The superconducting gap function $\Delta_{\alpha\beta}$ and 
the transition temperature are determined by
the self-consistent gap equation,
\begin{eqnarray}
\Delta_{\alpha\beta}=T\sum_{n,p}{\rm Tr}[\hat{\Gamma}^{\alpha\beta}(p,p')
\hat{F}(p')],
\end{eqnarray}
where $\hat{F}(p)$ is the anomalous Green function defined by eq. (\ref{fff}). 
We have introduced the four-point vertex function matrix
$\{\hat{\Gamma}^{\alpha\beta}(p,p')\}_{\gamma\delta}$.
This pairing interaction
consists of 
the parity-conserving part $\hat{\Gamma}_{\rm con}$ 
and the parity-nonconserving part $\hat{\Gamma}_{\rm noncon}$, i.e.,
$\hat{\Gamma}=\hat{\Gamma}_{\rm con}+\hat{\Gamma}_{\rm noncon}$.
The former can be decomposed into
 spin singlet and triplet channels.
Let us assume the following form of the pairing interaction.
\begin{eqnarray}
&&\{\hat{\Gamma}^{\alpha\beta}_{\rm con}(p,p')\}_{\gamma\delta}=
\Gamma^{(s)}_{\rm con}(p,p')
i(\sigma_{y})_{\alpha\beta}i(\sigma_{y})_{\gamma\delta} \nonumber \\
&&+\Gamma^{(t)}_{\rm con}(p,p')\mbox{\boldmath $\mathcal{L}$}(p)
\cdot (\mbox{\boldmath $\sigma$}
i\sigma_{y})_{\alpha\beta}\mbox{\boldmath $\mathcal{L}$}(p')
\cdot (\mbox{\boldmath $\sigma$}i\sigma_{y})_{\gamma\delta}.
\end{eqnarray}
The form of the triplet part is chosen to prevent
the pairing between the different Fermi surfaces due
to the mismatch between the $\mbox{\boldmath $d$}$-vector and
the momentum dependence of the SO interaction, as pointed out in the previous 
section.\cite{ede,fri,fuji2}
 The four-point vertices
$\Gamma^{(\nu)}_{\rm con}$ ($\nu=s,t$) are expanded 
in terms of the basis of the irreducible representations of
the point group of the system.
Generally, the parity-nonconserving part $\hat{\Gamma}_{\rm noncon}$ is 
nonzero, and expressed as
\begin{eqnarray}
&&\{\hat{\Gamma}^{\alpha\beta}_{\rm noncon}(p,p')\}_{\gamma\delta}
=\Gamma^{(st)}(p,p')i(\sigma_{y})_{\alpha\beta}
\mbox{\boldmath $\mathcal{L}$}(p')
\cdot (\mbox{\boldmath $\sigma$}i\sigma_{y})_{\gamma\delta} \nonumber \\
&& +\Gamma^{(ts)}(p,p')\mbox{\boldmath $\mathcal{L}$}(p)
\cdot (\mbox{\boldmath $\sigma$}
i\sigma_{y})_{\alpha\beta}i(\sigma_{y})_{\gamma\delta}.
\end{eqnarray}
$\Gamma^{(st)}$ and $\Gamma^{(ts)}$ are also classified in accordance with 
the irreducible representations.
The highest $T_c$ is achieved by  
$\Gamma^{(\nu)}(p,p')$ ($\nu=s,t,st,ts$)
which belongs to a unique irreducible representation, provided that
there is no accidental degeneracy.
Thus, $\Delta_{s}(p)$ and $\Delta_{t}(p)$ in eq.(\ref{delt}) are expressed by
the same basis function of the irreducible representation.
The superconducting state realized 
is the mixture of the spin singlet 
and triplet states.~\cite{ede,gr}
In this case, the triplet pairing state should be in a higher angular momentum
state than the singlet pairing state, and 
the possible pairing state is $s+p$ or $d+f$ or $g+h$, and so forth.

\begin{figure}
\begin{center}
\includegraphics[width=6cm]{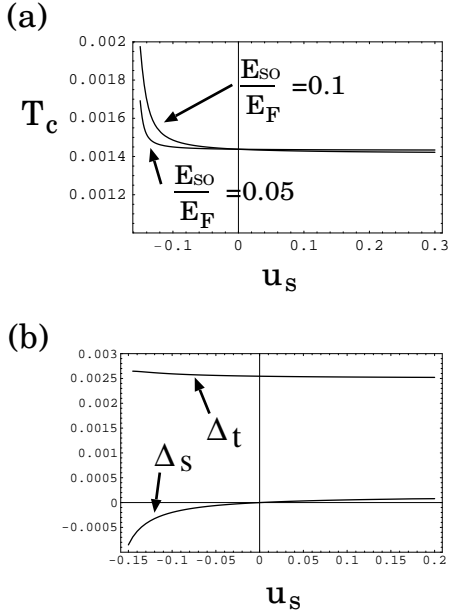}
\end{center}
\caption{(a) $T_c$ vs $u_s$. 
(b) $\Delta_s$ and $\Delta_t$ vs $u_s$ at zero temperature
for $E_{\rm SO}/E_F=0.1$.
}
\label{tcgap}
\end{figure}

To examine to what extent the admixture is realized, we carry out
a simple model calculation of the transition temperature $T_c$ and 
the gap function $\Delta$.
We consider a two-dimensional model with an isotropic Fermi surface and
the Rashba interaction $\mbox{\boldmath $\mathcal{L}$}_0(k)=(k_y,-k_x,0)$.
We also assume the existence of a pairing interaction in the $s$-wave 
and the $p$-wave channels expressed as
\begin{eqnarray}
V_{\rm pair}(k,k')
&=&u_sc^{\dagger}_{k\uparrow}c^{\dagger}_{-k\downarrow}c_{-k'\downarrow}c_{k'\uparrow}
\nonumber \\
&&+u_p[(i\hat{k}_{+})(i\hat{k}_{+}')c^{\dagger}_{k\uparrow}c^{\dagger}_{-k\uparrow}
c_{-k'\uparrow}c_{k'\uparrow} \nonumber \\
&&+(i\hat{k}_{-})(i\hat{k}_{-}')c^{\dagger}_{k\downarrow}c^{\dagger}_{-k\downarrow}
c_{-k'\downarrow}c_{k'\downarrow}], \label{pint}
\end{eqnarray} 
with $\hat{k}_{\pm}=(k_x\pm i k_y)/|\mbox{\boldmath $k$}|$,
and ignore any other interaction processes.
Then, the weak-coupling BCS gap equation is
\begin{eqnarray}
\Delta_{\tau}=\sum_{\tau'=\pm}g_{\tau\tau'}\sum_k
\Delta_{\tau'}\frac{\tanh (\frac{E_{k\tau'}}{2T})}{2E_{k\tau'}},
\quad (\tau=\pm) 
\end{eqnarray}
where $\Delta_{\tau}=\Delta_s+\tau\Delta_t$ with $\Delta_s$ and $\Delta_t$
the gap functions
for the spin singlet and spin triplet states, respectively. 
$g_{\tau\tau'}=-(u_s+\tau\tau' u_t)/2$ and
$E_{k\tau}=\sqrt{\varepsilon_{k\tau}^2+\Delta_{\tau}^2}$.
The transition temperature $T_c$ is easily calculated by using 
the linearized gap equation.
In Fig.\ref{tcgap}(a), we show the results of $T_c$
plotted as a function of $u_s$
for the attractive $p$-wave interaction with $u_p=-0.15 E_F$,
and $E_{\rm SO}/E_F=0.1$ or $E_{\rm SO}/E_F=0.05$
with $E_{\rm SO}$ the magnitude of the SO splitting.
For the attractive $s$-wave interaction $u_s<0$,
$T_c$ increases, as $|u_s|$ changes from 0 to $|u_p|$,
indicating that
the $s$-wave pairing mixes with the $p$-wave pairing cooperatively.
It is noted that for $u_s\approx u_p$, $T_c$ is substantially increased
by the admixture.
On the other hand, for the repulsive $s$-wave interaction $u_s>0$, 
$T_c$ is decreased by the admixture. The decrease in $T_c$ is at most 
of order $E_{\rm SO}/E_F$ for all positive values of $u_s$ including
an extreme case of $u_s/|u_p|\rightarrow \infty$.
To observe the admixture more directly, 
we show in Fig.\ref{tcgap}(b) the calculated results of 
the gap function for the spin triplet pair $\Delta_t$
and the spin singlet pair $\Delta_s$ in the case of $E_{\rm SO}/E_F=0.1$
at zero temperature.
For the attractive $u_s<0$, the mixed $s$-wave gap grows
significantly, as $|u_s|$ increases. The maximum ratio
of $\Delta_s/\Delta_t$ for $u_p<u_s<0$ is nearly equal to $0.32$.
In contrast to the substantial admixture for $u_s<0$,
$\Delta_s/\Delta_p$ is at most 
of order $E_{\rm SO}/E_F=0.1$ in the case of the repulsive $u_s>0$. 
We found that even in the limit of $u_s\rightarrow \infty$,
there exists a solution of the gap equation: $\Delta_t=0.002498$, 
$\Delta_s=0.0001468$. Thus, even when
there is a strong on-site Coulomb repulsion,
as in the case of heavy fermion systems,
the $s+p$ wave state may be possible, as long as
$E_{\rm SO}/E_F$ is sufficietly smaller than unity. 

In the above argument, it is postulated that
the $\mbox{\boldmath $d$}$-vector of the triplet pairing can be compatible 
with the momentum dependence of 
the SO interaction $\mbox{\boldmath $\mathcal{L}$}_0(k)$
to prevent the pairing between the different Fermi surfaces.
However, this condition is not generally guaranteed.
It may be possible that for some specific electronic structures,
the strongest attractive interaction in the triplet channel
has a momentum dependence inconsistent with 
$\mbox{\boldmath $\mathcal{L}$}_0(k)$, and thus 
the gap function $\hat{\Delta}(k)$ cannot be 
diagonalized by the unitary transformation 
$\hat{U}(k)\hat{\Delta}(k)\hat{U}^{t}(-k)$.
The realization of this pairing state 
depends on the competition between
the strength of the attractive interaction and
the pair-breaking effect caused by the pairing between
the different Fermi surfaces.

\section{Electromagnetic Properties}

\subsection{Paramagnetism}

The parity-breaking SO interaction yields unique paramagnetic properties.
For example, the transition between the two SO splitted Fermi surfaces
gives rise to
a van-Vleck-like spin susceptibility
in addition to the Pauli spin susceptibility.\cite{ede,gr,fri,fuji2}
The van-Vleck-like term is analogous to
the conventional van-Vleck orbital susceptibility.
However, there are some crucial differences between them,
as argued below.
An important feature of the van-Vleck-like term is
that when the SO splitting $E_{\rm SO}=\alpha |\mbox{\boldmath $\mathcal{L}$}|$
is much larger than the superconducting gap $\Delta$, as in the case
of CePt$_3$Si, CeRhSi$_3$, and CeIrSi$_3$,
this term is not strongly affected 
by superconducting transition,\cite{ede,gr,fri}
in contrast to the Pauli term which decreases below $T_c$ 
for spin singlet pairing states.

\begin{figure}
\begin{center}
\includegraphics[width=6cm]{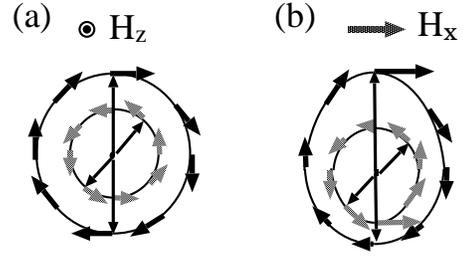}
\end{center}
\caption{Effects of the Zeeman magnetic field 
on the changes of the Fermi surfaces.
(a) For magnetic field parallel to $z$-axis.
Arrows on the Fermi surfaces indicate spins.
The length of the arrows represents the magnitude of magnetic moments. 
(b) For in-plane magnetic field.}
\label{f0}
\end{figure}

Here, we first consider the Rashba SO interaction relevant
to tetragonal systems, 
$\mbox{\boldmath $\mathcal{L}$}_0(k)=(t_{0y},-t_{0x},0)$, where
$t_{0\nu}$ ($\nu=x,y$) transforms like $k_{\nu}$.  
For small $|\mbox{\boldmath $k$}|$, $t_{0\mu}=k_{\mu}$.
In the case of CePt$_3$Si, CeRhSi$_3$, and CeIrSi$_3$,
the crystal structures of which
have $C_{4v}$ symmetry,
the $z$-component of $\mbox{\boldmath $\mathcal{L}$}_0$ is generally
nonzero, but given by $k_xk_yk_z(k_x^2-k_y^2)$ for small 
$|\mbox{\boldmath $k$}|$,
as elucidated by Samokhin.\cite{sam1}
However, to highlight the effects of the van-Vleck-like term, we exploit
the Rashba type interaction for simplicity.
From eq. (\ref{qene}), 
the single-particle energy in
the absence of the Coulomb interaction for $\mbox{\boldmath $H$}=(H_x,0,H_z)$ 
is given by
\begin{eqnarray}
\varepsilon_{k\pm}=\varepsilon_k\pm \alpha \sqrt{(t_{0y}-\mu_{\rm B}H_x)^2
+t_{0x}^2+\mu_{\rm B}H_z^2}.
\label{enera}
\end{eqnarray}
It is seen from eq. (\ref{enera}) that 
magnetic fields parallel to the $z$-axis merely change
the magnitude of the SO splitting of the Fermi surfaces, while
magnetic fields parallel to the $xy$-plane deform the Fermi surfaces
into asymmetric shapes, as depicted in Fig.\ref{f0}. 
These effects indicate that 
magnetic fields parallel to the $z$-axis is dominated by
the van-Vleck-like susceptibility, while
the in-plane magnetic fields induce responses from 
electron spins corresponding to
both the Pauli term and the van-Vleck-like term.

According to the Fermi liquid theory presented in section 2, 
the exact expression
of the spin susceptibility in the superconducting state 
for the Rashba interaction is obtained as\cite{fuji3}
\begin{eqnarray}
&&\chi_{zz}(T)=\mu_{\rm B}^2\sum_k\sum_{\tau=\pm}\tau z_{k\tau}
\frac{\tanh (\frac{E_{k\tau}}{2T})}{E_{k\tau}} \nonumber \\
&&\times\frac{E_{k\tau}^2+\varepsilon^{*}_{k+}\varepsilon^{*}_{k-}
+\Delta_{k+}\Delta_{k-}}{E_{k+}^2-E_{k-}^2} 
\Lambda^{sz}(E_{k\tau},\mbox{\boldmath $k$}), \label{scchizz}
\end{eqnarray}
for a magnetic field parallel to the $z$-axis, and
\begin{eqnarray}
&&\chi_{xx}(T)=\mu_{\rm B}^2\sum_{\tau=\pm}\sum_k\frac{z_{k\tau}}{4T
\cosh^2(\frac{E_{k\tau}}{2T})}
\hat{t}_y\Lambda_{\tau}^{sx}(E_{k\tau},\mbox{\boldmath $k$}) \nonumber \\
&&+\mu_{\rm B}^2\sum_k\sum_{\tau=\pm}
\tau z_{k\tau}
\frac{\tanh (\frac{E_{k\tau}}{2T})}{E_{k\tau}} \nonumber \\
&&\times\frac{E_{k\tau}^2+\varepsilon^{*}_{k+}\varepsilon^{*}_{k-}
+\Delta_{k+}\Delta_{k-}}{E_{k+}^2-E_{k-}^2} 
\hat{t}_x\Lambda^{sx}_{+-}(E_{k\tau},\mbox{\boldmath $k$}),
\label{scchi}
\end{eqnarray}
for an in-plane magnetic field.
Here $E_{k\tau}$ and $\Delta_{k\tau}$ are defined by the equations just
below eq. (\ref{scf})
The three-point vertices $\Lambda^{sz}$, $\Lambda^{sx}_{+-}$, and 
$\Lambda^{sx}_{\tau}$ are given by
\begin{eqnarray}
\Lambda^{sz}(p)=
1-\frac{1}{\mu_{\rm B}}\frac{\partial \Sigma_0(p)}{\partial H_z},
\label{svert1}
\end{eqnarray}
\begin{eqnarray}
\Lambda^{sx}_{\tau}(p)=\hat{t}_y
(1-\frac{1}{\mu_{\rm B}}\frac{\partial 
\Sigma_{x}}{\partial H_x})+\frac{\hat{t}_x}
{\mu_{\rm B}}\frac{\partial 
\Sigma_{y}}{\partial H_x}
-\frac{\tau}
{\mu_{\rm B}}\frac{\partial \Sigma_0}{\partial H_x},
\label{svert2}
\end{eqnarray}
\begin{eqnarray}
\Lambda_{+-}^{sx}(p)=\hat{t}_x(1-\frac{1}{\mu_{\rm B}}\frac{\partial 
\Sigma_{x}}{\partial H_x})
-\frac{\hat{t}_y}{\mu_{\rm B}}
\frac{\partial 
\Sigma_{y}}{\partial H_x}.
\label{svert3}
\end{eqnarray}
Here, $(\hat{t}_{x}(p),\hat{t}_{y}(p))=\hat{\mbox{\boldmath $t$}}(p)=
\mbox{\boldmath $t$}(p)/|\mbox{\boldmath $t$}(p)|$,
 $\mbox{\boldmath $t$}(p)=\mbox{\boldmath $n$}\times
\mbox{\boldmath {$\mathcal{L}$}}(p)$ with $\mbox{\boldmath $n$}=(001)$.
$\chi_{zz}(T)$ and the second term of $\chi_{xx}(T)$ are
the van-Vleck-like susceptibilities.
Both the Pauli and van-Vleck-like terms
are enhanced by electron correlation effects, because of
the three-point vertices. 

It is noted that the van-Vleck-like term is quite different from
the usual van-Vleck orbital susceptibility, which is
caused by symmetric SO interaction, for the following reason.
In contrast to the usual van-Vleck term, which is almost independent
of temperatures, the van-Vleck-like term
due to the parity-breaking SO interaction indeed depends
on temperatures, similarly to the Pauli susceptibility, since 
the magnitude of the parity-breaking SO splitting is much smaller
than the Fermi energy for typical heavy fermion systems.
According to a simple model calculation based on a two-dimensional 
tight-binding model on a square lattice carried out by the author, 
both the Pauli and van-Vleck-like terms exhibit strong temperature
dependence, as a result of the energy dependence of
the density of states.\cite{fuji3}
This property implies that it is almost impossible to distinguish
between these two contributions experimentally 
by analyzing the temperature dependence of spin susceptibilities.
Thus, we need to be careful about the interpretation
of experimental data of spin susceptibilities
in the superconducting state, since they always include 
the van-Vleck-like term which is not strongly affected by 
superconducting transition when 
$\Delta\ll E_{\rm SO}$.

In the case of the Rashba SO interaction, 
the temperature dependence of spin susceptibilities
below $T_c$ was calculated by Gorkov and Rashba, and 
Frigeri et al. for a model with
a spherical Fermi surface.\cite{gr,fri}
According to their results, 
for $\Delta\ll E_{\rm SO}$, 
$\chi_{xx}$ at zero temperature is one half of the magnitude 
in the normal state, i.e.,
 $\chi_{xx}(0)=\chi_{xx}(T_c)/2$, owing to the existence of
the van-Vleck-like term.

\begin{figure}
\includegraphics*[width=7cm]{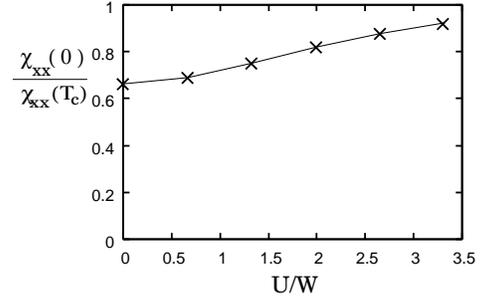}
\caption{\label{fig:eechira} $\chi_{xx}(T=0)/\chi_{xx}(T=T_c)$ 
in the superconducting state versus 
$U/W$ with $\alpha=1.0$ calculated 
by second order perturbative expansion with respect to $U$
for the 2D Hubbard-like model. 
} 
\end{figure}

However, experimental observations seem not to be consistent with
the theoretical prediction.
According to the Knight shift measurements performed 
by Yogi {\it et al.}\cite{yogi,yogi2} 
for CePt$_3$Si, 
both $\chi_{xx}$ and $\chi_{zz}$ 
show no significant change even below $T_c$. 
To explain the origin of the discrepancy, we note that
the magnitude of the ratio
$\chi_{xx}(T=0)/\chi_{xx}(T=T_c)$ in the superconducting state 
crucially depends on
the details of the electronic structure and electron correlation effects.
For instance, when the density of states has a strong energy dependence, 
$\chi_{xx}(T=0)/\chi_{xx}(T=T_c)$ may
deviate from 1/2.
To demonstrate this, we consider a Hubbard-type 
model on a square lattice with a kinetic energy 
$\varepsilon_k=-2(\cos k_x+\cos k_y)$, and the Rashba term with
$\mbox{\boldmath $\mathcal{L}$}_{0}(k)
=(\sin\frac{k_y}{2},-\sin\frac{k_x}{2},0)$.
For this model, at half-filling, 
we calculate $\chi_{xx}(0)/\chi_{xx}(T_c)$ in the case of 
$\Delta\ll \alpha |\mbox{\boldmath $\mathcal{L}$}|$,
handling electron correlation effects by perturbative expansion
in terms of $U$ up to the second order and using eq. (\ref{scchi}).
The results are shown in Fig. \ref{fig:eechira}
As $U/W$ increases, the van-Vleck-like susceptibility is more strongly
enhanced by
electron correlation
than the Pauli term, and the ratio
$\chi_{xx}(T=0)/\chi_{xx}(T=T_c)$ approaches 1.0.
Although the strength of the electron-electron interaction 
for which the ratio is nearly equal to 1.0
is considerably large, 
the results may be qualitatively valid, 
and can be improved by including higher order
corrections, as long as the system is in the Fermi liquid state.
As a matter of fact, 
CePt$_3$Si is in the Fermi liquid state with a mass enhancement factor
of order $\sim 60$, in which the spin susceptibility is also
substantially enhanced.
Thus, it is possible that, for CePt$_3$Si, 
because of strong electron correlation
and a particular electronic structure, $\chi_{xx}(T)$ is dominated by
the van-Vleck-like term, and thus is almost
independent of temperature below $T_c$, consistent with
the NMR measurements.

In the case with cubic symmetry, 
the uniform spin susceptibilities in the superconducting state
are obtained in a similar manner.\cite{fuji3}
The results are
\begin{eqnarray}
&&\chi_{zz}(T)=\mu_{\rm B}^2\sum_{\tau=\pm}\sum_k\frac{z_{k\tau}}{4T
\cosh^2(\frac{E_{k\tau}}{2T})}
\Lambda_{\rm P}^{\rm cub}(E_{k\tau},\mbox{\boldmath $k$}) 
\nonumber \\
&&+\mu_{\rm B}^2\sum_k\sum_{\tau=\pm}
\tau z_{k\tau}
\frac{\tanh (\frac{E_{k\tau}}{2T})}{E_{k\tau}} \nonumber \\
&&\times\frac{E_{k\tau}^2+\varepsilon^{*}_{k+}\varepsilon^{*}_{k-}
+\Delta_{k+}\Delta_{k-}}{E_{k+}^2-E_{k-}^2}
\Lambda^{\rm cub}_{\rm V}(E_{k\tau},\mbox{\boldmath $k$}), 
\label{sccu}
\end{eqnarray}
where 
\begin{eqnarray}
\Lambda^{\rm cub}_{\rm P}(p)&=&\hat{\mathcal{L}}_{0z}^2
(1-\frac{1}{\mu_{\rm B}}\frac{\partial \Sigma_0(p)}{\partial H_z})
\nonumber \\
&-&\frac{\hat{\mathcal{L}}_{0z}
\hat{\mathcal{L}}_x}{\mu_{\rm B}}
\frac{\partial \Sigma_x(p)}{\partial H_z}
-\frac{\hat{\mathcal{L}}_{0z}
\hat{\mathcal{L}}_y}{\mu_{\rm B}}
\frac{\partial \Sigma_y(p)}{\partial H_z},\label{tv1}
\end{eqnarray}
\begin{eqnarray}
\Lambda^{\rm cub}_{\rm V}(p)&=&(\hat{\mathcal{L}}_x^2
+\hat{\mathcal{L}}_y^2)
(1-\frac{1}{\mu_{\rm B}}\frac{\partial \Sigma_0(p)}{\partial H_z})
\nonumber \\
&+&\frac{\hat{\mathcal{L}}_{0z}
\hat{\mathcal{L}}_x}{\mu_{\rm B}}
\frac{\partial \Sigma_x(p)}{\partial H_z}
+\frac{\hat{\mathcal{L}}_{0z}
\hat{\mathcal{L}}_y}{\mu_{\rm B}}
\frac{\partial \Sigma_y(p)}{\partial H_z}.\label{tv2}
\end{eqnarray}
Because of the cubic symmetry, $\chi_{xx}=\chi_{yy}=\chi_{zz}$, and
$\chi_{zz}$ consists of both the Pauli term (the first term on
the right-hand side of eq. (\ref{sccu})) 
and van-Vleck-like term (the second term of (\ref{sccu})).
For systems with a spherical Fermi surface,
the Pauli term is 1/3 of the total susceptibility.
In this case, at zero temperature in the superconducting state, 
we have $\chi_{zz}(0)/\chi_{zz}(T_c)=2/3$ for $\Delta\ll E_{\rm SO}$.

\subsection{Pauli limiting field}

In this section, we consider the Pauli depairing effect due to
the Zeeman magnetic field.
This effect in the case without inversion symmetry was precisely
studied by Frigeri et al.\cite{fri} 
The changes of the Fermi surfaces due to the Zeeman magnetic fields  
depicted in Fig.\ref{f0} imply that
the Pauli depairing effect on Cooper pairs is also unique
in noncentrosymmetric superconductors.

We first consider the Rashba case with an asymmetric
potential gradient along the $z$-axis.
For magnetic fields parallel to the $z$-axis,
Cooper pairs between electrons with the momenta $k$ and $-k$ 
are always possible,
as shown in Fig.\ref{f0}(a), implying the suppression of the Pauli depairing 
effect.
More precisely, the Pauli limiting field depends on
the spin state of Cooper pairs.
For spin-triplet pairing states with the $\mbox{\boldmath $d$}$-vector
determined by the Rashba interaction, 
$\mbox{\boldmath $d$}\propto (t_{0y},-t_{0x},0)$,
the pairing state with
$S_z=\pm 1$ for the spin quantization axis parallel to the $z$-axis is
realized.
Magnetic fields parallel to the $z$-axis do not give rise
to the Pauli depairing effect on this state.
On the other hand, for spin singlet states,
the Pauli limiting field exists. However, it is strongly enhanced
by the parity-breaking SO interaction.
To argue the Pauli depairing effect on the singlet state,
we consider a simple case without electron correlation, and 
assume the attractive interaction $-Vw_kw_{-k}$ for a singlet channel,
where $w_k$ is a structure factor for the orbital degrees of freedom.
We also assume that $E_{\rm SO}/E_F$
is sufficiently small, and that the admixture of triplet states is negligible.
Then, the gap equation for $H_z\neq 0$ is given by
\begin{eqnarray}
\frac{1}{V}&
=&\sum_k\sum_{\tau=\pm 1}\frac{\alpha^2|t_0|^2w_k^2}{\alpha^2|t_0|^2
+\mu_{\rm B}^2H_z^2}\frac{\tanh \frac{E_{k\tau}}{2T}}{4E_{k\tau}} \nonumber \\
&+&\sum_k\sum_{\tau=\pm 1}\frac{w_k^2\mu_{\rm B}^2H_z^2}
{\alpha^2|t_0|^2+\mu_{\rm B}^2H_z^2}
\frac{\zeta_{k\tau}\tanh \frac{E_{k\tau}}{2T}}{4E_{k\tau}},
\label{mg}
\end{eqnarray}
where 
\begin{eqnarray}
\zeta_{k\tau}=1+\tau\frac{\alpha^2|t_0|^2+\mu_{\rm B}^2H_z^2}{\sqrt{\varepsilon_k^2
(\alpha^2|t_0|^2+\mu_{\rm B}^2H_z^2)+\mu_{\rm B}^2H_z^2\Delta_k^2}},
\end{eqnarray}
\begin{eqnarray}
&&E_{k\tau}=\pm [\varepsilon^2_k+\alpha^2|t_0|^2+ \mu_{\rm B}^2H_z^2
+\Delta_k^2 \nonumber \\
&&+2\tau \sqrt{\alpha^2|t_0|^2+\mu_{\rm B}^2H_z^2}
\sqrt{\varepsilon_k^2+\Delta_k^2(\xi^2_{+}-\xi^2_{-})^2}
]^{\frac{1}{2}},
\end{eqnarray}
and  $|t_0|^2=t_{0x}^2+t_{0y}^2$, 
$\Delta_k=\Delta_0w_k$.
It is evident that for $\mu_{\rm B}H_z\ll \alpha |t_0|$,
the first term on the right-hand side of eq. (\ref{mg}) dominates
over the second term, and thus the Pauli depairing effect
is negligible.
For general values of $H_z$,
the transition temperature $T_c$ is determined by using the linearized gap
equation.
It is noted that the first term on the right-hand side of (\ref{mg})
in the linearized approximation 
yields a $\log T$ singularity 
at low temperatures for any finite values of $H_z$ which satisfies
$\mu_{\rm B}H_z<E_F$, while
the second term exhibits no logarithmic singularity.
As a result, the finite transition temperature $T_c$ always exists
for any finite magnetic field $H_z< E_F$.
This implies that the Pauli limiting field is infinitely large for
sufficiently low temperatures.
According to precise numerical calculations performed by Frigeri et al.,
the Pauli limiting field determined by the linearized gap equation
obeys a concave curve on the $H$-$T$ plane at sufficiently 
low temperatures.\cite{fri} 
It is noteworthy that the recent experiments on
CeRhSi$_3$ support the existence of
the large upper critical fields at low temperatures
exceeding a conventional Pauli limit, and exhibiting
a concave behavior as a function of $T$.\cite{kim3}
In these experiments, although the concave behavior is observed even
near $T_c$, it is expected that the inclusion of the strong-coupling effect 
enlarges the temperature region where the concave behavior appears
up to higher temperatures close to $T_c$.

Also, the above results are easily obtained from
a simple thermodynamic argument.
Let $\chi_{zz,N}$ and $\chi_{zz,S}$ be the spin susceptibilities in the
normal and superconducting states, respectively,
for a magnetic field parallel to the $z$-axis.
Then, the Pauli limiting field $H_P$ at zero temperature is determined by
equating the ground state energy in the normal state to 
that in the superconducting state
\begin{eqnarray}
\frac{\chi_{zz,N}H_P^2}{2}=\frac{\chi_{zz,S}H_P^2}{2}+\frac{N(0)\Delta^2_0}{2}, 
\end{eqnarray}
which leads to 
$H_P=\Delta_0\sqrt{N(0)/(\chi_{zz,N}-\chi_{zz,S})}$,
where $N(0)$ is the average density of states
of the SO splitted two bands.
We see from eq.(\ref{scchizz}) that 
when $\Delta_0\ll E_{\rm SO}$, 
$\chi_{zz,N}-\chi_{zz,S}\approx\chi_{zz,N}(\Delta_0/E_{\rm SO})^2\ll \chi_{zz,N}$
holds, leading to a large $H_P$ enhanced by the factor
$E_{\rm SO}/\Delta_0$.
More precisely, the free energy difference between the superconducting state
and the normal state at any finite temperatures is calculated from
$\Omega_s-\Omega_n=\int_0^{\Delta_0}(d V^{-1}/d\Delta)\Delta^2d\Delta$.
By carrying out numerical calculations,
one can confirm that when the gap equation (\ref{mg}) 
gives a nonzero value of 
$\Delta_0$,
$\Omega_s-\Omega_n<0$ holds at least for $\Delta_0 < E_{\rm SO}$, 
and thus the superconducting state
is the absolute minimum of the free energy;
i.e. $H_P$ determined by using the linearized gap equation coincides
with that obtained from the thermodynamic consideration.

For magnetic fields parallel to the $xy$-plane, 
the Fermi surfaces are deformed into asymmetric shapes as shown in
Fig. \ref{f0}(b), which gives rise to the Pauli depairing effect 
for both spin singlet and spin triplet states.
The Pauli depairing effect stems from
the Pauli term of the spin susceptibility $\chi_{xx}$, i.e.,
the first term of eq. (\ref{scchi}).  
When $\Delta_0\ll E_{\rm SO}$, and 
$\chi^{\rm van ~Vleck}_{xx,N}-\chi_{xx,S}^{\rm van ~Vleck}
\approx \chi_{xx,N}^{\rm van ~Vleck}
(\Delta_0/E_{\rm SO})^2\ll \chi_{xx,N}^{\rm van ~Vleck}$, $\chi_{xx,N}^{\rm Pauli}$
holds,
the Pauli limiting field is 
$H_P\approx\Delta_0\sqrt{N(0)/(\chi_{xx,N}^{\rm Pauli}-\chi_{xx,S}^{\rm Pauli})}$.
In the case with a spherical Fermi surface but without 
electron correlation, 
$\chi_{xx,N}^{\rm Pauli}$ is one half of the total spin
susceptibility. Thus, at zero temperature where $\chi^{\rm Pauli}_{xx,S}=0$, 
$H_P$ is enhanced by a factor of $\sqrt{2}$, compared to the case
with inversion symmetry in which the Pauli limiting field is
$H_P^{\rm inv.sym.}=\Delta_0/\sqrt{2}$.
However, the real situation is more complicated.
According to a more precise analysis
done by Kaur et al. and Samokhin, 
the asymmetric deformation of the Fermi surface due to
in-plane magnetic fields may stabilize
Cooper pairs with a center of mass momentum, 
leading to the Fulde-Ferrel state.\cite{kau,sam2}
We would like to discuss this point in section 4.6.

In the case with cubic symmetry,
the spin susceptibility consists of both the Pauli and
van Vleck-like terms, and an argument similar to
that for the Rashba model with in-plane magnetic fields
is applicable. When $\Delta_0\ll E_{\rm SO}$ holds, 
the Pauli limiting field is given by
$H_P\approx\Delta_0\sqrt{N(0)/(\chi_{zz,N}^{\rm Pauli}-\chi_{zz,S}^{\rm Pauli})}$.
Here, $\chi_{zz,S(N)}^{\rm Pauli}$ is the first term on the right-hand side of
eq. (\ref{sccu}) in the superconducting (normal) state.
In the case with a spherical Fermi surface without electron correlation,
at zero temperature, $\chi^{\rm Pauli}_{zz,N}$ is 1/3 of the total
susceptibility, as claimed in the previous section, and thus 
$H_P$ is enhanced by a factor of
$\sqrt{3}$ compared with that in the case with inversion symmetry.

\subsection{Nuclear relaxation rate $1/T_1$}

One of powerful experimental probes for
superconducting states is 
the NMR measurement of the nuclear relaxation rate $1/T_1$.
For conventional superconductors with the $s$-wave symmetry,
$1/T_1$ exhibit a coherence peak just below $T_c$, and 
an exponential decay at low temperatures.
For unconventional superconductors with an inversion center
such as High-$T_c$ cuprates, and many heavy fermion systems,
there is no coherence peak, because of the absence of the coherence factor
and the suppressed singularity of the density of states, and also
a power law decay at low temperatures appears. 
In the case without inversion symmetry, a quite different behavior
is theoretically expected, as pointed out by the present author and 
Hayashi et al.\cite{fuji2,haya}

For simplicity, we neglect electron correlation effects,
which may not change the following argument drastically. 
Then, the nuclear relaxation rate is given by
\begin{eqnarray}
&&\frac{1}{T_1T}\propto  \nonumber \\
&&\lim_{\omega\rightarrow 0}
\frac{1}{\omega}{\rm Im}\Bigl[T\sum_{\varepsilon_m}\sum_{k,k'}
\{{\rm Tr}[\frac{\sigma^{+}}{2}\hat{G}(\varepsilon_m+\omega_n,k)
\frac{\sigma^{-}}{2}\hat{G}(\varepsilon_m,k')] \nonumber \\
&&-{\rm Tr}[\frac{\sigma^{+}}{2}
\hat{F}(\varepsilon_m+\omega_n,k)
\frac{\sigma^{-}}{2}\hat{F}(\varepsilon_m,k')]\}
|_{i\omega_n\rightarrow \omega+i\delta}\Bigr] \nonumber \\
&&=\int \frac{d\varepsilon}{2\pi}\frac{1}{2T\cosh^2\frac{\varepsilon}{2T}}
\{[N_n(\varepsilon)]^2+[N_a(\varepsilon)^2]\}, \label{t1}
\end{eqnarray}
with $N_n(\varepsilon)$ and $N_a(\varepsilon)$ defined by the retarded Green's
functions as
\begin{eqnarray}
N_n(\varepsilon)=-\sum_k\sum_{\tau=\pm}{\rm Im}
G^R_{\tau}(\varepsilon,k),
\end{eqnarray}
\begin{eqnarray}
N_a(\varepsilon)=-\sum_k\sum_{\tau=\pm}{\rm Im}
F^R_{\tau}(\varepsilon,k). \label{cf}
\end{eqnarray}
$N_a(\varepsilon)$ gives the coherence factor, which enhances
the coherence peak of $1/T_1T$.
For unconventional pairing states with the angular momentum $\ell \geq 2$,
the coherence factor vanishes.
A remarkable feature of $1/T_1T$ for 
noncentrosymmetric superconductors appears
in the case of the $p$-wave state.
In contrast to the case with inversion symmetry in which 
the coherence factor vanishes for the $p$-wave state,
the absence of inversion symmetry gives rise to
nonzero coherence factor $N_a$ for the $p$-wave state.
This is easily understood as follows.
In the case of the pure $p$-wave state, the superconducting gap is given by
$\hat{\Delta(k)}=\Delta_t\mbox{\boldmath 
$\mathcal{L}$}(k)\cdot\mbox{\boldmath $\sigma$}i\sigma_y$ where
$\mbox{\boldmath $\mathcal{L}$}(k)$ has the $p$-wave symmetry.
An important point is that the factor $\mbox{\boldmath $\mathcal{L}$}(k)$
appears in eq. (\ref{cf}) only in the form 
$|\mbox{\boldmath $\mathcal{L}$}(k)|$. 
(see eqs.(\ref{deldia}) and (\ref{scf}))
Thus, when the density of states for the $(+)$-band
is different from that for the $(-)$-band, 
$N_a(\varepsilon)$ is nonzero.
This result implies that the coherence peak of $1/T_1T$ just below $T_c$
is markedly enhanced in the case of the $p$-wave pairing dominated
state. 
The implication of this result for CePt$_3$Si will be discussed in
section 5.

%

\subsection{Magnetoelectric effect in the normal state}

The existence of the asymmetric SO interaction 
$\alpha\mbox{\boldmath $\mathcal{L}$}_0(k)
\cdot\mbox{\boldmath $\sigma$}$ yields a nontrivial coupling between
charge and spin degrees of freedom, giving rise to
interesting electromagnetic properties.
One example is the existence of magnetoelectric effects.
Magnetoelectric effects in a normal metal were discussed
by Levitov et al. many years ago within one-body approximation.\cite{lev,ede0}

\begin{figure}
\begin{center}
\includegraphics[width=8cm]{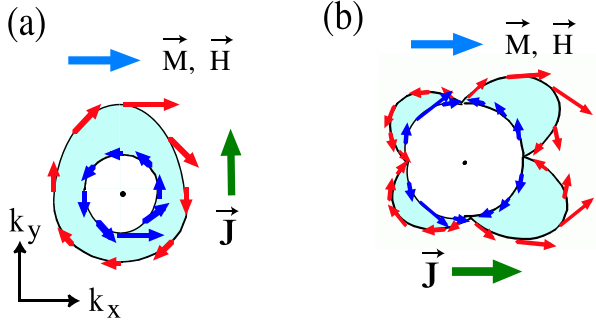}
\end{center}
\caption{(Color online)
Changes of the Fermi surfaces and the spin configuration 
due to applied magnetic field (or net current flow).
(a) For the Rashba interaction, current flow along the $y$-axis gives rise to
magnetization along the $x$-axis. Conversely, an applied magnetic field
along the $x$-axis yields current flow along the $y$-axis.
(b) For the Dresselhaus interaction, the current flow is parallel to
the applied magnetic field (or induced magnetization). A sectional view for 
$k_z=0$ is shown.}
\label{f1}
\end{figure}

In the case of the Rashba SO interaction with 
$\mbox{\boldmath $\mathcal{L}$}_0(k)=(t_{0y}(k),-t_{0x}(k),0)$,
the magnetization along the $x$-direction is generated
by an electric field applied in the $y$-direction as 
\begin{eqnarray}
M_x=\Upsilon_{xy}E_y,
\end{eqnarray}
and conversely
an AC-magnetic field gives rise to a charge current flow expressed as
\begin{eqnarray}
J_x=-2\Upsilon_{xy}\frac{\partial B_y}{\partial t}.\label{jb}
\end{eqnarray}
The physical origin of these effects is easily understood as follows.
As shown in Fig. \ref{f1} (a),
when charge current flows along the $y$-axis,
the Fermi surface is deformed, and because of the Rashba SO interaction,
the distribution of spins changes, yielding the bulk magnetization
along the $x$-axis.
Conversely, an applied magnetic field in the $x$-direction changes
the distribution of spins, which also deforms the Fermi surface asymmetrically,
leading to a charge current in the $y$-direction.
    
In the case of heavy fermion systems, the magnetoelectric effect
coefficient $\Upsilon_{xy}$ is significantly 
affected by strong electron correlation.
According to the analysis based on the Fermi liquid theory,
we obtain a simple relation among
$\Upsilon_{xy}$, the specific heat coefficient $\gamma$, and the resistivity
$\rho$:\cite{fuji3}
\begin{eqnarray}
\Upsilon_{xy}\sim \frac{\mu_{\rm B}}{e v_F^{*}\rho}
\cdot\frac{\alpha k_F}{E_F} \propto\frac{\gamma}{\rho}.
\label{ups}
\end{eqnarray} 
In general, for heavy fermion systems, 
the resistivity is given by $\rho\sim \rho_0+{\rm const.}\gamma^2T^2$,
with $\rho_0$ a residual resistivity.
Thus, at sufficiently low temperatures, $\Upsilon_{xy}$ is enhanced by
the factor $\gamma$.

We now estimate the order of the magnitude of these effects.
We assume that the Fermi velocity is $v_{F}^{*}\sim 10^5$ cm/s, 
which corresponds to a mass enhancement of order $\sim 100$, and 
the SO splitting is sufficiently large, e.g., $\alpha k_F/E_F\sim 0.1$.
To consider the magnetization induced by an electric field,
we assume that 
the charge current density is $J\sim 1 ~\mbox{A/cm$^2$}$.
Then, the induced magnetization is estimated as,
$M=\Upsilon_{xy}E_y\approx\mu_{\rm B}(\alpha k_F/E_F)(J/ev_{F}^{*})\sim 1$ Gauss,
which is experimentally measurable.
To evaluate the charge current induced by AC magnetic fields 
we assume that an AC magnetic field $B=B_0\cos(\omega t)$ with
$B_0\sim 100 $ Gauss, and $\omega\sim100$ kHz is applied, and that
the normal resistivity is $\rho\sim 10 \mu\Omega\cdot{\rm cm}$.
Then we obtain the charge current,
$J=-2\Upsilon_{xy}(dB/dt)\approx \mu_{\rm B}(\alpha k_F/E_F)(dB/dt)/(ev_F^{*}\rho)
\sim 1$ mA/cm$^2$. This magnitude is also experimentally accessible.
However, in this case, it is necessary to distinguish between
the magnetoelectric effect and the usual eddy current induced by
the time-dependent magnetic field.
To avoid the eddy current, leads attached to the sample must be
aligned exactly parallel to the magnetic field. 

A similar effect is also possible in cubic systems without inversion symmetry.
In this case, the charge current (or applied electric fields)
is parallel to magnetic fields (or magnetization) as shown in 
Fig.\ref{f1} (b), 
i.e., $\mbox{\boldmath $J$}=-2\Upsilon (d\mbox{\boldmath $B$}/dt)$.
The coefficient $\Upsilon$ is approximately given by
eq. (\ref{ups}).

\subsection{Magnetoelectric effect in the superconducting state}

Magnetoelectric effects in the superconducting state
analogous to those in the normal state are possible.
These phenomena were theoretically
predicted by Edelstein, and extensively discussed by Yip and
the present author.\cite{ede,ede2,yip,yip2,fuji2,fuji3}
In the absence of inversion symmetry,
supercurrent can be induced by the Zeeman magnetic field,
which may be called ``paramagnetic supercurrent''.\cite{ede,yip} 
Conversely, bulk magnetization is raised by supercurrent 
flow.\cite{ede2}
The physical origin of these effects is also
the asymmetric deformation of the Fermi surface due to an applied magnetic
field as in the case of the normal state. 
However, in contrast to the analogous effects in the normal state,
the magnetoelectric effects in the superconducting state
are dissipationless. As a result, to detect the paramagnetic
supercurrent flow experimentally, one needs to prepare a system in which
the dissipationless bulk current is possible.
One example of such a system may be realized by attaching leads made
of superconductors to the sample.\cite{gl} 

To explain the above-mentioned effects in more detail, 
we first use the Ginzburg-Landau (GL) theory,
and later, we will present microscopic analysis.
The GL free energy for superconductors without inversion symmetry
was derived by Edelstein, Samokhin, and Kaur et al.\cite{ede3,sam2,kau},
which reads,
\begin{eqnarray}
&&F_s-F_n=a|\Psi|^2
+\frac{\beta}{2}|\Psi|^4+\frac{1}{2m_{\mu}}|D_{\mu}\Psi|^2 \nonumber
\\
&&+\frac{\mathcal{K}_{\mu\nu}}{2en_s}B_{\mu}
(\Psi (D_{\nu}\Psi)^{*}+\Psi^{*} D_{\nu}\Psi) \nonumber \\
&&+\frac{\mbox{\boldmath $B$}^2}{8\pi}
-\frac{\chi_{\mu\mu} B_{\mu}^2}{2},
\label{freene}
\end{eqnarray}
where $a=a_0(T-T_{c0})$, $D_{\mu}=-\hbar\nabla_{\mu}-2eA_{\mu}/c$, 
$A_{\mu}$ is a vector potential,
$\mbox{\boldmath $B$}=\nabla\times \mbox{\boldmath $A$}$,  
$\mbox{\boldmath $M$}$ is the magnetization density, and
$n_s$ is the superfluid density.
The fourth term of eq. (\ref{freene}) 
with the coefficient $\mathcal{K}_{\mu\nu}$
stems from the parity-breaking SO interaction, and is the origin
of magnetoelectric effects.
Differentiating the free energy with respect to $\mbox{\boldmath $A$}$ 
and $\mbox{\boldmath $B$}$, we obtain the following relations 
for the supercurrent density $\mbox{\boldmath J}^s$ and the magnetization 
density $\mbox{\boldmath $M$}$.
\begin{eqnarray}
J_{\mu}^s=J^{\rm dia}_{\mu}
+\mathcal{K}_{\nu\mu}B_{\nu}, \label{scab}
\end{eqnarray}
\begin{eqnarray}
M_{\mu}=\mathcal{K}_{\mu\nu}\Lambda J^{\rm dia}_{\nu}
+M_{\mu}^{\rm Zee}.
\label{mab}
\end{eqnarray}
Here, $J^{\rm dia}_{\mu}$ is the conventional diamagnetic supercurrent
given by $J^{\rm dia}_{\mu}=(\hbar\nabla_{\mu}\phi-2e A_{\mu}/c)/(2e\Lambda)$.
$\phi$ is the phase of the order parameter $\Psi$.
$\Lambda^{-1}=2e^2|\Psi|^2/m$.
$M^{\rm Zee}_{\mu}$ is a magnetization due to the conventional Zeeman effect.
Also, we have put $|\Psi|^2=n_s$.
The second term on the right-hand side of eq. (\ref{scab}) 
is the paramagnetic supercurrent, and the 
first term on the right-hand side of eq. (\ref{mab})
is the magnetization induced by supercurrent flow.
In the case of the Rashba interaction, $\mathcal{K}_{\mu\nu}$ is nonzero
only for $(\mu,\nu)=(x,y)$ or $(y,x)$.
In the case of cubic systems, $\mathcal{K}_{\mu,\nu}=\mathcal{K}\delta_{\mu\nu}$.

As was pointed out by Yip, in the case of the Rashba interaction,
the paramagnetic supercurrent is partially canceled with
the magnetization current 
$\mbox{\boldmath $J$}^M=c\nabla\times\mbox{\boldmath $M$}$.\cite{yip2}
To observe this, using eqs.(\ref{scab}), (\ref{mab}), and the relation 
$\nabla\times \mbox{\boldmath $J$}^{\rm dia}=-\mbox{\boldmath $B$}/c\Lambda$,
we write down the total current,
\begin{eqnarray}
&&\mbox{\boldmath $J$}_s+\mbox{\boldmath $J$}_M 
=\mbox{\boldmath $J$}^{\rm dia}+c\nabla\times\mbox{\boldmath $M$}_{\rm Zee}
\nonumber \\
&&+c\mathcal{K}\Lambda(-\partial_xJ_z^{\rm dia},\partial_yJ_z^{\rm dia},
\partial_xJ_x^{\rm dia}+\partial_yJ_y^{\rm dia}). \label{jt}
\end{eqnarray}
The last term on the right-hand side of eq. (\ref{jt})
is the paramagnetic supercurrent.
In the complete Meissner state and in the thermodynamic limit, 
this term vanishes, and thus there is no paramagnetic supercurrent.
Yip pointed out that because of this cancellation, 
the penetration depth is symmetric under the transformation 
$z\rightarrow -z$.\cite{yip2}
However, in finite systems, or in the mixed state, the last term
of eq. (\ref{jt}) gives
nonzero contributions to the magnetoelectric effect. 

In the case of cubic systems, the cancellation between the paramagnetic
supercurrent and the magnetization current is perfect.
This is easily seen as follows.
Since in cubic systems $\mathcal{K}_{\mu\nu}=\mathcal{K}\delta_{\mu\nu}$, 
we obtain
$\mbox{\boldmath $J$}^s=\mbox{\boldmath $J$}^{\rm dia}
+\mathcal{K}\mbox{\boldmath $B$}$,  
$\mbox{\boldmath $M$}=\mathcal{K}\Lambda
 \mbox{\boldmath $J$}^{\rm dia}+\mbox{\boldmath $M$}^{\rm Zee}$, 
and $\mbox{\boldmath $J$}^M=c\nabla\times\mbox{\boldmath $M$}
=-\mathcal{K}\mbox{\boldmath $B$}+c\nabla\times\mbox{\boldmath $M$}^{\rm Zee}$,
and as a result, 
the paramagnetic current $\mathcal{K}\mbox{\boldmath $B$}$ cancels
exactly in the total current $\mbox{\boldmath $J$}^s+\mbox{\boldmath $J$}^M$.
There  is  no paramagnetic supercurrent in this case.

The magnetoelectric effect coefficient $\mathcal{K}_{\mu\nu}$
can be obtained from microscopic calculations as in the case of
the normal state.
According to the Fermi liquid theory explained in section 2, 
the formula of $\mathcal{K}_{yx}$, 
which includes electron correlation effects exactly, is 
given by\cite{fuji2,fuji3} 
\begin{eqnarray}
&&\frac{\mathcal{K}_{yx}}{e\mu_{\rm B}}= \nonumber \\
&&\sum_k\sum_{\tau=\pm 1}\tau v_{0y\tau}
\frac{z_{k\tau}\Delta_{k\tau}^2}{E_{k\tau}^{2}}
\left[\frac{{\rm ch}^{-2}\frac{E_{k\tau}}{2T}}{2T}-
\frac{{\rm th}\frac{E_{k\tau}}{2T}}{E_{k\tau}}\right]
\Lambda^{sx}_{\tau}(E_{k\tau},\mbox{\boldmath $k$}) 
\nonumber \\
&&+2\alpha\sum_k\frac{\Delta_{k+}\Delta_{k-}}
{E_{k+}^2-E_{k-}^2}\left[z_{k-}\frac{{\rm th}\frac{E_{k+}}{2T}}{E_{k+}}
-z_{k+}\frac{{\rm th}\frac{E_{k-}}{2T}}{E_{k-}}\right] \nonumber \\
&&\times
\hat{t}_x\Lambda^{sx}_{+-}(E_{k\tau},\mbox{\boldmath $k$}), \label{kyx}
\end{eqnarray}
where $v_{0y\tau}=\partial (\varepsilon_k
+\tau\alpha |t_0|)/\partial k_{y}$.
In the case with a spherical Fermi surface, up to the first order
in $\alpha k_F/E_F$, the magnetoelectric coefficient is simplified as
\begin{eqnarray}
\mathcal{K}_{\mu\nu}=\frac{e\mu_{\rm B}n_s\alpha}{8\pi^3zE_F},
\end{eqnarray}
where $n_s$ is the superfluid density.
Note that $\mathcal{K}_{yx}$ is amplified by 
the mass enhancement factor $1/z_{k\tau}$,
when the Wilson ratio is nearly equal to 2, as realized in
typical heavy fermion systems.
This feature is in contrast to the electron correlation effect on
the diamagnetic supercurrent which is
suppressed by the factor $z_{k\tau}$.
As a result, the magnetoelectric effect in the superconducting state
is more strongly enhanced 
in heavy fermion systems with large effective mass
than in weakly correlated metals.

We now discuss the feasibility of experimental observations of these effects.
Assuming that $\alpha k_F/E_F\sim 0.1$, 
the electron density $n\sim 10^{22} ~{\rm cm}^{-3}$, 
the mass enhancement factor $1/z\sim 100$, and 
$v_s/v_F^{*}\sim \Delta/E_F\sim 0.01$, we estimate
the magnitude of the bulk magnetization induced by the supercurrent as
$M=\mathcal{K}\Lambda J^{\rm dia}\approx 
\mu_{\rm B}n(\alpha k_F/E_F)(v_s/v_F^{*})/(8\pi^3 z)\sim 0.1$ Gauss.
The experimental detection of this internal field may be possible.
Under the above conditions,
the magnitude of the paramagnetic supercurrent 
is also accessible by
conventional experimental measurements.
It should be emphasized again that 
to detect paramagnetic supercurrent,
one needs to prepare a circuit in which 
bulk current flow without dissipation is possible.
Also, in the mixed state, vortices should be pinned by impurities
to suppress dissipation due to the flux flow, which may be induced by
supercurrent, as demonstrated by Oka et al.\cite{ichi}

\subsection{Helical vortex state (Fulde-Ferrel state)}

In the previous section, we have seen that 
the asymmetric deformation of the Fermi surface
due to an applied magnetic field may give rise to
the paramagnetic supercurrent, if the dissipationless bulk current
flow is permitted.
Conversely, when bulk current flow is forbidden, as in the case of
an isolated system, the asymmetric deformation of the Fermi surface
stabilizes an inhomogeneous superconducting state with 
a finite phase gradient.
This possibility was first
pointed out by Kaur et al. and Samokhin.\cite{kau,sam2}
They argued that in the Rashba case,
an in-plane magnetic field applied to an isolated system 
gives rise to a Fulde-Ferrel-like state,
in which Cooper pairs have a center of mass momentum and the phase of
the order parameter is spatially modulated.\cite{kau,sam2}
Kaur et al. called this novel phase the helical vortex state.
Following their analysis,
we would like to consider this phase on the basis of 
the GL free energy (\ref{freene}). From the derivative of
eq. (\ref{freene}) with respect to $\Psi$, one finds that
a stable solution is given by
$\Psi(R)=\Psi_0(R)e^{-i\mbox{\boldmath $q$}\cdot\mbox{\boldmath $R$}}$
with\cite{kau}
\begin{eqnarray}
q_{\mu}=-\frac{2m\mathcal{K}_{\nu\mu}}{\hbar en_s}B_{\nu}.
\end{eqnarray}
Here $\Psi_0(R)$ is the solution of the GL free energy 
without the parity-breaking term (the fourth term of eq.(\ref{freene})).
This corresponds to a Fulde-Ferrel state with a spatial modulation of
the order parameter phase.
It is noted that the period of the modulation $1/q_{\mu}$ decreases 
as $\sim z_{k\tau}^2$, as the effective electron mass $1/z_{k\tau}$ increases.
When $1/q_{\mu}$ is smaller than the spacing between vortices,
the helical vortex phase governs the upper critical field.
The critical temperature at finite $\mbox{\boldmath $B$}$ in this phase is
derived by using the standard method,\cite{kau,sam2}
\begin{eqnarray}
T_c(\mbox{\boldmath $B$})=T_c(0)-\frac{\pi B}{\Phi_0m a_0}
+\frac{\mathcal{K}_{\mu\nu}^2m}{8a_0 e^2 n_s^2}B^2. \label{metc}
\end{eqnarray}
Thus, the transition temperature (or the upper critical field)  
is increased by the inversion-symmetry
breaking term of the free energy (\ref{freene}).
We would like to emphasize that the increase in $T_c(\mbox{\boldmath $B$})$
is drastically amplified by electron correlation effects, since 
the last term of the right-hand side of (\ref{metc})
is enhanced by the factor $1/z^3_{k\tau}$, 
while the second term corresponding to
the conventional orbital depairing effect 
is suppressed by the factor $z_{k\tau}$.
Therefore, the realization of the helical vortex state 
is feasible in heavy fermion systems.
To estimate the order of the magnitude of this effect,
let us assume that, as in the case of CePt$_3$Si,   
the ratio of the SO splitting to the Fermi energy $\alpha k_F/E_F$
is $\sim 0.1$, 
the mass enhancement factor $1/z_{k\tau}$ is of order $\sim 100$,
the Fermi velocity is $v_F^{*}\sim 10^{5}$ cm/s, 
the upper critical field is $H_{c2}\sim 4$ T, and the applied magnetic field $B$
is close to $H_{c2}$.
Then, the period of the spatial modulation is estimated as
$1/q\sim \mu_{\rm B}B/(16\pi^3v_F^{*}z)\cdot(\alpha k_F/E_F)\sim 10^{-6}$ cm,
which is compatible with the inter-vortex distance.
In this situation, 
the increase in $T_c$ due to the Helical vortex phase is
almost on the same order as the decrease due to
the orbital depairing effect, and plays a crucial role
in the vicinity of the upper critical field.

\subsection{Anomalous Hall effect}

\begin{figure}
\begin{center}
\includegraphics[width=6cm]{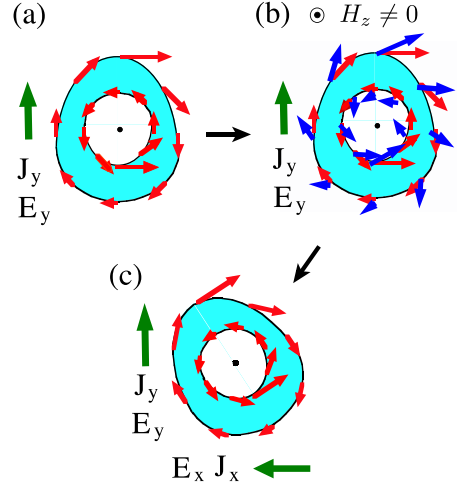}
\end{center}
\caption{(Color online) Schematic explanation for anomalous Hall effect.
(a) An applied electric field along the $y$-axis changes
the distribution of spin. (b) Torque produced by 
magnetic field along the $z$-axis rotates spins. 
(c) Because of the Rashba SO interaction, the Fermi surface is deformed,
leading to a transverse current along the $x$-axis. 
}
\label{f2}
\end{figure}

In general,
the SO interaction can yield
the anomalous Hall effect, as clarified by Karplus-Luttinger many years 
ago.\cite{kl}
The anomalous Hall effect stems from the anomalous velocity which
is caused by the SO interaction.
In the case of the Rashba SO interaction,  
this mechanism can be schematically illustrated as shown
in Fig.\ref{f2}.
When an electric field applied along the $y$-axis induces
current flow along this direction, the Fermi surfaces
are deformed asymmetrically, and thus
the distribution of spins, which is constrained by the Rashba SO interaction,
is changed as shown in Fig.\ref{f2}(a).
Then, a magnetic field applied along the $z$-axis gives rise to
a torque which rotates spins around the $z$-axis as depicted in Fig.\ref{f2}(b).
Since the Rashba SO interaction forces the Fermi momentum 
to be perpendicular to spins, the asymmetric Fermi surfaces
are rotated on the $xy$-plane. 
As a consequence, the net current along
the $x$-axis occurs as shown in Fig.\ref{f2}(c). 
In the case of the Rashba interaction,
the anomalous velocity
$\mbox{\boldmath $v$}_{\rm A}=\alpha\nabla \mbox{\boldmath $t$}_0(k)
\cdot(\mbox{\boldmath $n$}
\times\mbox{\boldmath $\sigma$})$
is perpendicular to the $z$-axis.
Thus, the anomalous Hall effect is possible only for magnetic fields
along the $z$-axis. 
According to the analysis based on the Fermi liquid theory,
the anomalous Hall conductivity in the normal state 
is expressed as\cite{fuji3} 
\begin{eqnarray}
\frac{{\rm Re}~\sigma_{xy}^{\rm AHE}}{H_z}&=&
e^2\mu_{\rm B}\sum_{\tau=\pm}\sum_k
\frac{-\tau f(\varepsilon^{*}_{k\tau})
\Lambda^{sz}(
\varepsilon_{k\tau}^{*},\mbox{\boldmath $k$})}
{2\alpha|{\rm Re}~\mbox{\boldmath $t$}
(\varepsilon_{k\tau}^{*},\mbox{\boldmath $k$})|^3} \nonumber \\
&&\times (\partial_{k_{x}}t_{x\tau}\partial_{k_{y}}t_{0y}
-\partial_{k_{x}}t_{y\tau}\partial_{k_{y}}t_{0x}). \label{ahe}
\end{eqnarray}
This quantity is enhanced by the factor $\Lambda^{sz}$, which 
is equal to the enhancement factor of the van-Vleck-like
spin susceptibility (\ref{scchizz}).
For heavy fermion systems, this factor is of the same order as
the mass enhancement factor $1/z_{k\tau}\approx 100\sim 1000$, and
thus, the anomalous Hall effect is significantly large.
For instance, let us assume  
the resistivity $\rho\sim 10 ~\mu\Omega\cdot {\rm cm}$,
the mass enhancement factor $1/z_{k\tau}\sim 100$, 
the Fermi velocity $v_{F}^{*}\sim 10^5$ cm/s,
and the carrier density $n\sim 10^{22}$ cm$^{-3}$. Then,
the ratio of $\sigma^{\rm AHE}_{xy}\sim e^2\mu_{\rm B}B/(h^2v_F^{*}z)$ 
to the normal Hall conductivity 
$\sigma^{\rm NHE}_{xy}$ is estimated as 
$\sigma^{\rm AHE}_{xy}/\sigma^{NHE}_{xy}\sim 40$.
The anomalous Hall effect dominates over the normal Hall effect.

An analogous effect of heat current is also possible.
The anomalous Hall conductivity for heat current
is expressed as\cite{fuji3}
\begin{eqnarray}
\kappa_{xy}^{\rm AHE}=\frac{1}{T}(L^{(2)}_{xy}
-\sum_{\mu\nu}L^{(1)}_{x\mu}L^{(0)-1}_{\mu\nu}L^{(1)}_{\nu y}), \label{hh}
\end{eqnarray}
where $L^{(0)}_{\mu\nu}$ is equal to the conductivity tensor
$\sigma_{\mu\nu}$ and
\begin{eqnarray}
&&\frac{L_{xy}^{(m) {\rm AHE}}}{H_z}=e^{2-m}
\mu_{\rm B}\sum_{\tau=\pm}\sum_k(-\tau)
(\varepsilon_{k\tau}^{*})^m f(\varepsilon^{*}_{k\tau})
\nonumber \\
&&\times
\frac{
\Lambda^{sz}(\varepsilon_{k\tau}^{*},
\mbox{\boldmath $k$})}
{2\alpha|{\rm Re}~\mbox{\boldmath $t$}
(\varepsilon_{k\tau}^{*},\mbox{\boldmath $k$})|^3} 
(\partial_{k_{x}}t_{x\tau}\partial_{k_{y}}t_{0y}
-\partial_{k_{x}}t_{y\tau}\partial_{k_{y}}t_{0x}), \label{tahll}
\end{eqnarray}
with $m=1,2$.

Note that in eqs. (\ref{ahe}) and (\ref{tahll}), 
electrons away from the Fermi surface give dominant contributions
to the anomalous Hall conductivity.
This feature is in accordance with the fact that
the magnetic response against the magnetic field along the $z$-axis
is governed by the van-Vleck-like term.
This observation leads us to an interesting implication for
the superconducting state.
In the superconducting state, the Hall effect for heat current 
is possible at finite temperatures,
and when
the superconducting gap is much smaller than the size of the SO splitting, 
the thermal anomalous Hall conductivity is not affected by
the superconducting transition.
Furthermore, even in the limit of zero temperature,  
$\kappa_{xy}^{\rm AHE}/(TH_z)$ is nonzero, 
and behaves as in the normal state, 
even though the quasiparticle density is vanishingly small.
The experimental detection of this effect is an intriguing future issue.

\section{Pairing State Realized in CePt$_3$Si}

In recent years, several heavy fermion superconductors without inversion
symmetry have been discovered.
In particular, extensive experimental studies on the superconducting state
in CePt$_3$Si have revealed several important aspects of this system.
In this section, we would like to discuss about possible pairing states
realized in CePt$_3$Si, comparing the recent experimental data
with the theoretical understanding of noncentrosymmetric
superconductors.

As mentioned in the previous sections,
the $\mbox{\boldmath $d$}$-vector of the spin triplet component
is determined by the parity-breaking SO interaction.
In the case of CePt$_3$Si which has the crystal structure with
the $C_{4v}$ symmetry, the possible form of the parity-breaking SO
interaction $\alpha\mbox{\boldmath $\mathcal{L}$}_0(k)
\cdot \mbox{\boldmath $\sigma$}$ 
was clarified by Samokhin from a general group theoretical
argument.\cite{sam1}
According to his analysis, 
there are three constraint conditions that determine 
the form of $\mbox{\boldmath $\mathcal{L}$}_0(k)$:
(i) $\mbox{\boldmath $\mathcal{L}$}_0(-k)=
-\mbox{\boldmath $\mathcal{L}$}_0(k)$.
(ii) $\mbox{\boldmath $\mathcal{L}$}_0(k)$ satisfies
the symmetry of the point group.
(iii) $\mbox{\boldmath $\mathcal{L}$}_0(k)$ 
is a pseudovector which changes its sign under the transformation
$(k_x,k_y,k_z)\rightarrow (k_y,k_x,k_z)$,
$(x,y,z)\rightarrow (y,x,z)$, since
it is a quantum mechanical average of the operator 
$\nabla V\times \mbox{\boldmath $k$}$.\cite{sam1,com}
Then, the general form of
$\mbox{\boldmath $\mathcal{L}$}_0(k)$ for $C_{4v}$ symmetry is
given by
$\mbox{\boldmath $\mathcal{L}$}_0(k)
=(k_y,-k_x,c_0k_xk_yk_z(k_x^2-k_y^2))$.\cite{sam1}
For periodic systems, $k_{\mu}$ and $k_{\mu}^2$ are, respectively,
replaced by $\sin k_{\mu}$ and $\cos k_{\mu}$.
The coefficient $c_0$ of the $z$-component depends on microscopic 
detail of the system.
The case of $c_0=0$ corresponds to the Rashba interaction.
In this case, the $\mbox{\boldmath $d$}$-vector proportional to
$(k_y,-k_x,0)$ gives the highest $T_c$, as clarified by Frigeri et al.\cite{fri}
In the case of $c_0\neq 0$, the situation is more complicated, because
the magnitude of 
the pairing interaction for the channel corresponding to the basis function
$k_xk_yk_z(k_x^2-k_y^2)$ (the $A_2$ representation of $C_{4v}$) is generally
different from that for the $p$-wave channel corresponding to
the $E$ representation of $C_{4v}$, and
thus $\mbox{\boldmath $d$}\neq 
\Delta_t(k)\mbox{\boldmath $\mathcal{L}$}_0(k)$,
leading to depairing effects, as mentioned in section 2.
The pairing state in this case is determined by the competition between
the strength of the pairing interaction and the pair-breaking effect
due to the SO interaction.
In the following, we assume that $c_0$ is sufficiently small for simplicity,
and that the Rashba type interaction is applicable to this system, i.e.,
$\mbox{\boldmath $d$}\propto (k_y,-k_x,0)$.

%

\begin{figure}
\begin{center}
\includegraphics[width=5cm]{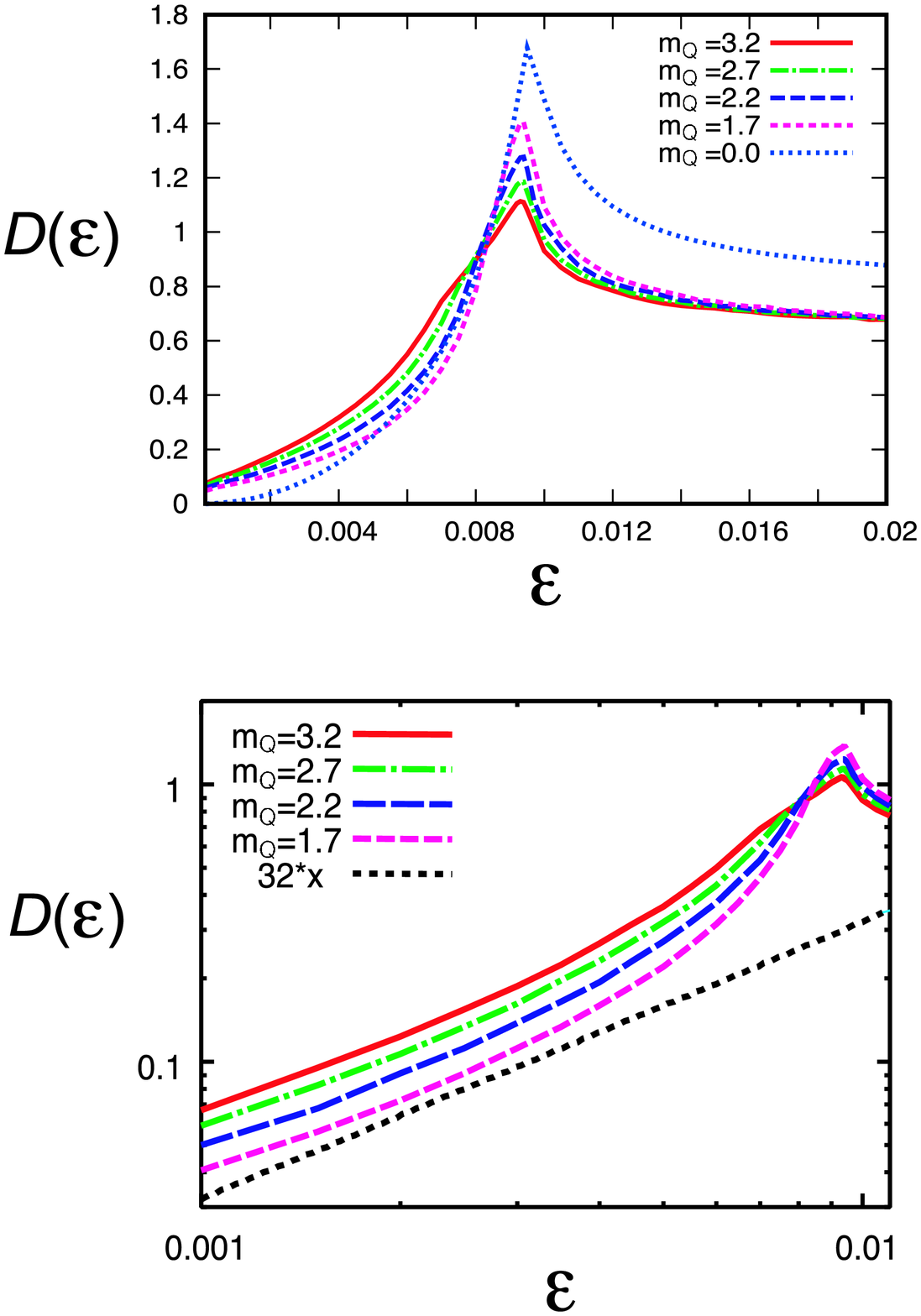}
\end{center}
\caption{(Color online) Density of states 
for $p$-wave pairing state plotted against 
excitation energy $\varepsilon$. The upper panel is
on a linear scale. The lower panel is a log-log plot.}
\label{f4}
\end{figure}

\begin{figure}
\begin{center}
\includegraphics[width=2.7cm]{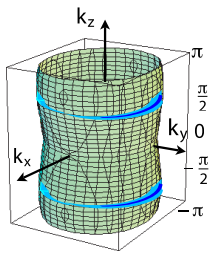}
\end{center}
\caption{(Color online) 
Schematic figure of line node structure on the Fermi surface.
The depth of the blue color in the vicinity of the magnetic Brillouin zone
$k_z\pm\pi/2$ indicates the depth of the node of the excitation gap.}
\label{f6}
\end{figure}

One remarkable experimental observation for CePt$_3$Si 
which is relevant to the pairing state
was obtained from the NMR measurements
of $1/T_1$ by Yogi et al.\cite{yogi,yogi2}
According to their data, $1/T_1$ exhibit a small coherence peak
just below $T_c$, indicating a fully gapped state.
Another important experimental observation was obtained from
the thermal transport measurements by Izawa et al.\cite{iza}
They found that at low temperatures the thermal conductivity exhibits
the Volovik effect, indicating
the existence of line nodes of the superconducting gap.
Also, the results of
the penetration depth measurement,\cite{pene}
and the specific heat measurement\cite{on}
support the existence of the line nodes
at sufficiently low temperatures.
Thus, for the clarification of the pairing state,
it is crucial to reconcile the existence of the coherence peak of $1/T_1$
with the line node of the superconducting gap.
In general, the coherence peak of $1/T_1$ is due to the coherence factor 
and the singularity of the density of states at the gap edge.
Even for unconventional superconductors with line nodes,
the singularity of the density of states exists, and may gives rise to
a small coherence peak.
However, in strongly correlated electron systems, because of
the large quasiparticle damping, the singularity is smeared out,
and as a result, no coherence peak is observed for all unconventional
superconductors in heavy fermion systems except CePt$_3$Si.
In CePt$_3$Si, the quasiparticle damping is quite large as indicated by 
the resistivity data in the normal state.
Thus, the existence of a coherence peak cannot be explained by
the singularity of the density of states.
Also, it is very unlikely that the $s$-wave pairing is realized 
in such heavy fermion systems with the large onsite repulsion.
Therefore, we need to give some specific reason for this remarkable
feature.
As argued in section 4.3, in noncentrosymmetric superconductors,
the coherence peak of $1/T_1$ is enhanced even for the $p$-wave pairing
dominated states, because of the non-vanishing coherence factor.
Thus, from the argument presented in section 3, 
one possible candidate for the pairing state realized 
in CePt$_3$Si is the $s+p$ wave state, in which 
the $p$-wave pairing is predominant.
The present author and Hayashi et al., independently 
proposed this pairing state.
Furthermore, Hayashi et al. pointed out
that in the $s+p$ state, when the magnitude of the $s$-wave gap $\Delta_s$
is close to that of the $p$-wave gap $\Delta_p$,
a line node of the excitation gap appears, because 
$\Delta_{-}=\Delta_s-\Delta_p\sin\theta$ can be zero for
a specific $\theta$, where $\theta$ is the azimuthal angle.
Hayashi et al. attributed the line nodes observed at low temperatures
in CePt$_3$Si to this mechanism.\cite{haya}
This scenario is tempting, since all the experimental observations
are explained only by tuning one parameter $\Delta_s/\Delta_p$.
However, as claimed above, the substantial weight of the $s$-wave admixture
is a bit unlikely in heavy fermion systems in which
there is a strong onsite repulsion.  
The present author proposed another scenario in which 
the coupling with an antiferromagnetic (AF) order that coexists with 
the superconductivity in CePt$_3$Si 
gives rise to nodal structures of the excitation gap.\cite{fuji4}
The drastic change in the excitation spectrum due to
the coexistence of the AF order is not inherent in noncentrosymmetric
superconductors, but was pointed out for centrosymmetric conventional
superconductors previously.\cite{mac}
In the case of CePt$_3$Si,
the AF order is characterized by the ordering $Q$-vector along 
the $(001)$ direction and
the staggered moment parallel to the $ab$-plane.\cite{neu}
We demonstrated that for sufficiently large magnitudes of the staggered moment,
a line node structure appears in the vicinity of the magnetic Brillouin zone
even when the superconducting gap itself has no nodes.
Assuming the $p$-wave pairing state coexisting with
the AF order, we calculated
the energy dependence of the density of states $D(\varepsilon)$.
The results are shown in Fig. \ref{f4}.
Here, $m_Q$ is the magnitude of the staggered moment.
In the calculation, we assumed the Rashba type SO interaction, and
the $\mbox{\boldmath $d$}$-vector is $\mbox{\boldmath $d$}=(k_y,-k_x,0)$.
At sufficiently low energies, $D(\varepsilon)\propto \varepsilon$ holds;
a characteristic of line nodes. 
In this calculation, it is assumed that the staggered moment is aligned 
parallel to the (100) direction. 
In Fig. \ref{f6}, we show a schematic figure of the line node
structure of the Fermi surface.
As a matter of fact, this node structure is not a true node, but
a minimum of the excitation gap. Thus, the coherence factor
is not suppressed by this node structure, which is
in accordance with the existence of the coherence peak of $1/T_1$.
It is noted that the node structure
has  $C_{2v}$ symmetry in momentum space because 
the direction of the staggered moment breaks $C_{4v}$ symmetry.
The experimental detection of this two-fold symmetry
is a crucial test for this scenario.

The possibility of the realization of the $s+p$ wave state 
should be examined by microscopic calculations.
For CePt$_3$Si that has a tetragonal crystal structure with $C_{4v}$ symmetry,
basis functions for the $s$-wave states
corresponding to the irreducible representations for $A_{1g}$
are $1$, $\cos k_x+\cos k_y$, and $\cos k_z$,
as clarified by Sergienko and Curnoe.\cite{ser}
Thus, a possible candidate for the gap function  is
given by eq.(\ref{delt}) with 
$\Delta_s(k)=\Delta^{(0)}_s+\Delta^{(1)}_s(\cos k_x+\cos k_y)
+\Delta^{(2)}_s\cos k_z$, $\Delta_t(k)=\Delta^{(0)}_t+\Delta^{(1)}_t
(\cos k_x+\cos k_y)+\Delta^{(2)}_t \cos k_z$,
and the Rashba type $\mbox{\boldmath $\mathcal{L}$}(k)=(\sin k_y, -\sin k_x,0)$.
Even for this gap function, since the isotropic $s$-wave component 
$\Delta^{(0)}_s$
is generally nonzero, and gives rise to the coupling with
the strong repulsive onsite interaction,
its stability should be examined by model calculations.
Unfortunately, it is difficult to construct
a reliable microscopic model for CePt$_3$Si at the present moment, since
experimental studies of 
microscopic electronic structure such as
the de Haas-van Alphen effect have not yet succeeded in detecting
the Fermi surface of heavy electrons in CePt$_3$Si.\cite{dh}
Very recently, Yanase and Sigrist have carried out microscopic calculations 
using the Fermi surface obtained by the LDA calculation, and found
that the admixture of a p-wave state and an extended s-wave state
is stabilized.\cite{yana}

\section{Concluding Remarks}

In this review, we survey our theoretical understanding of 
the unique properties of noncentrosymmetric superconductors.
One of the main conclusions is that
the remarkable effects due to parity-violation
are strongly enhanced by electron correlation, and thus
may play crucial roles in heavy fermion systems. From
this point of view, the recently discovered superconductors without
inversion symmetry such as CePt$_3$Si, CeRhSi$_3$, CeIrSi$_3$, and UIr
are quite fascinating systems.
It is highly desirable to reveal the interesting features
inherent in noncentrosymmetric superconductors
experimentally in these novel systems in the near future.

We should note that this review, by no means, encompasses all topics 
on noncentrosymmetric superconductors.
In particular, we could not present arguments on 
issues on the mixed state in detail.
Semiclassical approaches to these subjects have been   
explored by Hayashi et al. and Nagai et al.\cite{sem,kato}.
In the case of $\Delta/E_{\rm SO} \ll 1$ as realized in all heavy fermion
superconductors without inversion symmetry, the pairing states
on different Fermi surfaces can be approximately separated 
even under applied magnetic fields, and thus the theoretical treatment is
much simplified.
However, when $\Delta/E_{\rm SO}\sim 1$, with finite magnetic fields,
the pairing between the different Fermi surfaces can not be negligible,
and presumably yields drastic changes of low-energy properties.
As a matter of fact, as pointed out recently by Eremin and Annett,
the Zeeman field destroys 
the line node structure due to the admixture of the $s$-wave pairing and
the $p$-wave pairing considered by Hayashi et al.\cite{haya} and mentioned
in section 5, leading to
a full-gap state.\cite{ane}  
The theoretical clarification of orbital effects in such cases
is an interesting future issue.

\acknowledgments{}

The author is grateful to K. Yamada, Y. Onuki, M. Sigrist, S. K. Yip,
D. Agterberg,  
Y. Matsuda, T. Shibauchi, N. Kimura, T. Takeuchi, R. Settai, 
M. Ichioka, Y. Yanase, H. Mukuda, M. Yogi, 
and H. Ikeda for valuable discussions.
This work was partly supported by a Grant-in-Aid from the Ministry
of Education, Culture, Sports, Science and Technology, Japan.


\begin{thebibliography}{9}
\bibitem{com} More precisely, the symmetry of 
$\mbox{\boldmath $\mathcal{L}$}_0(k)$ for general $\mbox{\boldmath $k$}$
away from the $\Gamma$ point is lower than the point group symmetry, and thus
$\mbox{\boldmath $\mathcal{L}$}_0(k)$ for general $\mbox{\boldmath $k$}$
should be determined only by microscopic first principle calculations
such as the LDA.
\end{thebibliography}
\end{document}